\documentclass[dvipsnames, 11pt]{article}
\usepackage[utf8]{inputenc}

\usepackage{jheppub}

\usepackage{hyperref}
\usepackage{amsmath}
\usepackage{amssymb}
\usepackage{tikz}
\usetikzlibrary{calc}

\usepackage[toc,page]{appendix}
\usepackage{enumerate}
\usepackage{tensor}
\usepackage{booktabs}
\usepackage{bbm}
\usepackage{youngtab}
\usepackage{slashed}
\usepackage{multirow}
\usepackage{makecell}
\usepackage{float}
\usepackage{comment}
\usepackage{verbatim}
\usepackage{xcolor}
\usepackage[framemethod=tikz]{mdframed}
\usepackage{soul}
\usepackage{mathtools}

\usepackage{empheq}

\newcommand{\ab}{|}

\newcommand{\de}{\mathrm{d}}

\newcommand{\e}{\mathrm{e}}
\newcommand{\I}{\mathrm{i}}

\newcommand{\NS}{{\mathrm{NS}}}
\newcommand{\R}{{\mathrm{R}}}

\newcommand{\be}{\begin{equation}}
\newcommand{\ee}{\end{equation}}
\newcommand{\ba}{\begin{eqarray}}
\newcommand{\ea}{\end{eqarray}}

\newcommand{\N}{{\cal N}}

\counterwithin{table}{section}

\allowdisplaybreaks

\hypersetup{
    colorlinks=true,
    linkcolor=Maroon,
    citecolor=Maroon,
    filecolor=Maroon,      
    urlcolor=Maroon,
}

\usepackage{titlesec} 
\titleformat{\paragraph}
{\normalfont \normalsize \itshape}{\theparagraph}{2ex}{}
\titlespacing*{\paragraph}
{0pt}{3.25ex plus 1ex minus .2ex}{1.5ex plus .2ex}

\usepackage{chngcntr}

\numberwithin{equation}{section}

\title{Modular invariance, misalignment and finiteness in non-supersymmetric strings}

\author[a,b]{Niccol\`o Cribiori,}
\author[c]{Susha Parameswaran,}
\author[c]{Flavio Tonioni,}
\author[d,a]{and Timm Wrase}

\affiliation[a]{Institute for Theoretical Physics, TU Wien,\\
Wiedner Hauptstrasse 8-10/136, A-1040 Vienna, Austria}
\affiliation[b]{Max-Planck-Institut f\"ur Physik (Werner-Heisenberg-Institut),\\
F\"ohringer Ring 6, 80805, M\"unchen, Germany}
\affiliation[c]{Department of Mathematical Sciences, University of Liverpool,\\
Mathematical Sciences Building, Liverpool, L69 7ZL, UK}
\affiliation[d]{Department of Physics, Lehigh University,\\
16 Memorial Drive East, Bethlehem, PA 18018, USA}

\emailAdd{cribiori@mpp.mpg.de}
\emailAdd{susha@liv.ac.uk}
\emailAdd{flavio.tonioni@liv.ac.uk}
\emailAdd{timm.wrase@lehigh.edu}

\abstract{
In this article we show that finite perturbative corrections in non-supersymmetric strings can be understood via an interplay between modular invariance and misaligned supersymmetry. While modular invariance is known to be crucial in closed-string models, its presence and role for open strings is more subtle. Nevertheless, we argue that it leads to cancellations in physical quantities such as the one-loop cosmological constant and prevents them from diverging. In particular, we show that if the sector-averaged number of states does not grow exponentially, as predicted by misaligned supersymmetry, all exponential divergences in the one-loop cosmological constant cancel out as well. To account for the absence of power-law divergences, instead, we need to resort to the modular structure of the partition function. We finally comment on the presence of misaligned supersymmetry in the known 10-dimensional tachyon-free non-supersymmetric string theories.
}

\notoc

\begin{document}

\begin{flushright}
MPP-2021-179
\end{flushright}

\maketitle

\newpage

\section{Introduction} \label{sec: introduction}

The mathematical structure underlying superstring theory has received a tremendous and well-deserved amount of attention. However, it is a fact that the real world is not supersymmetric (at least at low energies), and there are reasons to believe that non-supersymmetric string-theory models should possess as much mathematical elegance as their supersymmetric counterparts. In fact, even if in these models there is no supersymmetry, it has been proposed that there is often nevertheless a special pattern in the bosonic and fermionic degrees of freedom, whereby the typical spectrum exhibits an increasing oscillation between net-bosonic and net-fermionic state degeneracies at each level.  This has been called `misaligned supersymmetry' \cite{Dienes:1994np, Dienes:1995pm, Dienes:2001se} and it is the subject of the present work.  More on the mathematical side, in ref.~\cite{Angelantonj:2010ic} an intriguing connection to the Riemann hypothesis has been proposed, thus pointing towards the presence of a rich and interesting structure behind non-supersymmetric models.

The original formulation of misaligned supersymmetry involved only closed strings \cite{Dienes:1994np, Dienes:2001se}. Recently, we showed explicitly in ref.~\cite{Cribiori:2020sct} that misaligned supersymmetry is also present in certain open-string models with broken supersymmetry in which anti-D$p$-branes are placed on top of O$p$-planes (see also ref.~\cite{Niarchos:2000kw} for related work involving open strings). This is consistent with the conjecture in ref.~\cite{Israel:2007nj}, which states essentially that non-supersymmetric theories without open strings, like the heterotic string theories, have to admit misaligned supersymmetry, while, in theories with oriented and unoriented closed and open strings, the open-string sector needs to have misaligned supersymmetry, but not necessarily the closed-string sector.

Misaligned supersymmetry has been used to provide a heuristic explanation as to how string theory is capable of giving finite answers even without supersymmetry \cite{Dienes:1994np, Dienes:2001se}.
A given non-supersymmetric string theory consists of a number of different sectors, each with their own infinite tower of physical states. The net boson-fermion degeneracies, $a_n^i$ in each sector $i$, can be computed from the corresponding partition-function characters via a Hardy-Ramanujan-Rademacher expansion, and they each grow exponentially as $a_n^i \sim e^{C\sqrt{n}}$. Despite this exponential growth in the mismatch between bosons and fermions, Kutasov and Seiberg showed in ref.~\cite{Kutasov:1990sv} that, in non-supersymmetric string theories that are modular-invariant and free of physical tachyons, an \emph{asymptotic supersymmetry} is observed in the high energy limit, leading to a cancellation between bosons and fermions. In ref.~\cite{Dienes:1994np}, Dienes observed that non-supersymmetric closed oriented string theories actually exhibit the oscillating pattern of \emph{misaligned supersymmetry} at all energy levels. So, bosons and fermions never cancel at any given level but rather exhibit an oscillation between net bosonic and net fermionic states. Dienes defined a sector-averaged net degeneracy $\langle a_n \rangle$, and proved that for oriented closed-string theories, the exponential growth in $\langle a_n \rangle$ is always slower than the growth in the individual sectors, $\langle a_n \rangle \sim e^{C_{\mathrm{eff}}\sqrt{n}}$ with $C_{\mathrm{eff}}<C$, provided modular invariance and the absence of physical tachyons. Dienes moreover conjectured that all exponential growth in the sector-average cancelled, $C_{\mathrm{eff}}=0$, leaving only polynomial growth with $n$. These remarkable cancellations in the sector-averaged degeneracies provided a way to characterise the finiteness of non-supersymmetric string theories with misaligned spectra. In ref.~\cite{Cribiori:2020sct}, we extended the results to open-string setups and proved the conjecture that all exponential growth is cancelled in an appropriately defined sector-average.  However, there was no clear, direct relationship between the sector-average and finite physical observables.

The purpose of this paper is to make these heuristic arguments precise, and to provide an explicit proof of how a misaligned spectrum ensures cancellations in physical quantities and leads to finite results. This represents a physically intuitive explanation that should parallel the usual argument based on modular invariance. In particular, we show that the cancellations that take place in the heuristic sector-averaged net degeneracy also appear directly in the one-loop cosmological constant. We find indeed that misalignment leads to a cancellation of all exponential divergences in the latter, and we expect a similar structure to emerge in the other quantum-corrected observables too. The modular properties of the partition functions further lead leftover power-law divergences to cancel, leading to an overall finite result. Although the role of modular invariance in ensuring finiteness is well-known for closed-string theories, we prove that similar cancellations also hold for open strings. This might be considered surprising, as modular invariance is explicitly broken by the worldsheet boundary. In this respect, we will argue that a remnant of the original modular group is enough to explain finiteness in the open-string models we analyze.

As is well known, string theory provides a huge multitude of vacua, which makes it hard to draw general conclusions. For this reason, in this article we restrict our analysis to 10-dimensional models, prior to any compactification. There are only a small number of known 10-dimensional consistent superstring models \cite{Mourad:2017rrl}. Of course, there are the five supersymmetric and anomaly-free consistent theories: type IIA and type IIB theories, which are closed-string theories with $\N_{10}=2$ supersymmetries, heterotic $\mathrm{E}_8 \times \mathrm{E}_8$- and $\mathrm{SO}(32)$-theories, that are also closed-string theories but with $\N_{10}=1$ supersymmetry, and type I $\mathrm{SO}(32)$-theory, which involves both closed and open strings and has $\N_{10}=1$ supersymmetry. Moreover, there also exist three non-supersymmetric, tachyon-free theories:
\begin{itemize}
    \item the heterotic $\mathrm{SO}(16) \!\times\! \mathrm{SO}(16)$-theory \cite{AlvarezGaume:1986jb, Dixon:1986iz}, with the eponymous gauge group;
    \item the Sugimoto model \cite{Sugimoto:1999tx}, with a gauge group $\mathrm{USp}(32)$;
    \item the type $0^\prime$B theory \cite{Sagnotti:1995ga, Sagnotti:1996qj, Seiberg:1986by}, with a gauge group $\mathrm{SU(32)}$.
\end{itemize}
A key difference of the Sugimoto model with the other non-supersymmetric models is the presence of a gravitino in its spectrum. The absence of a Lagrangian mass term for the gravitino is still compatible with supersymmetry breaking at the string scale, which leads indeed to a non-linear realisation in the spacetime effective theory \cite{Dudas:2000nv}. This is the simplest instance of a scenario that goes by the name of `brane supersymmetry breaking' \cite{Sugimoto:1999tx,Antoniadis:1999xk,Angelantonj:1999jh,Aldazabal:1999jr,Angelantonj:1999ms,Dudas:2000ff,Dudas:2000nv,Pradisi:2001yv,Blumenhagen:1998uf,Blumenhagen:1999ns}. While the heterotic $\mathrm{SO}(16) \!\times\! \mathrm{SO}(16)$-theory and the Sugimoto model have their entire spectrum in a standardly or misalignedly supersymmetric phase, and therefore do exhibit misaligned supersymmetry, this is not the case for the type $0^\prime$B theory. The latter is somewhat special since it presents misaligned supersymmetry only in the open-string sector (annulus and Möbius strip), whereas its closed-string sector does not present any sort of supersymmetry whatsoever, containing only bosons. Nevertheless, the closed-string sector is not tachyonic and therefore it has no divergence in physical quantities like the cosmological constant. It turns out that the structure ensuring the absence of UV-divergences in the latter can also be described in terms of a misaligned action of the orientifold symmetry in the Klein bottle, which alternatively adds or removes states from the halved torus at integer and semi-integer levels. The result is a closed-string spectrum which, albeit being purely bosonic, exhibits an oscillating growth of the number of states with the energy.

This paper is organized as follows. In section \ref{sec: misSUSY review} we briefly review the core ideas behind misaligned supersymmetry. In section \ref{sec: Lambda}, we review the relationship between the one-loop cosmological constant and the partition function in string theory. In section \ref{sec: openstring-misSUSY}, we show the details of how misaligned supersymmetry guarantees a finite one-loop cosmological constant for open strings and then in section \ref{sec: closedstring-misSUSY} we discuss the same topic for closed strings. In section \ref{sec: supertraces}, we provide an interpretation of open-string supertraces. In section \ref{sec: allnonSUSYmodels}, we comment on the presence of misaligned supersymmetry in the known 10-dimensional non-supersymmetric theories. We summarize the main results of the article in section \ref{sec: conclusions}. After this, appendix \ref{app:specialfunctions} reviews useful properties of special functions appearing in misaligned supersymmetry and appendix \ref{app: bl-derivation} contains additional computational details.

\section{Misaligned supersymmetry: a review} \label{sec: misSUSY review}
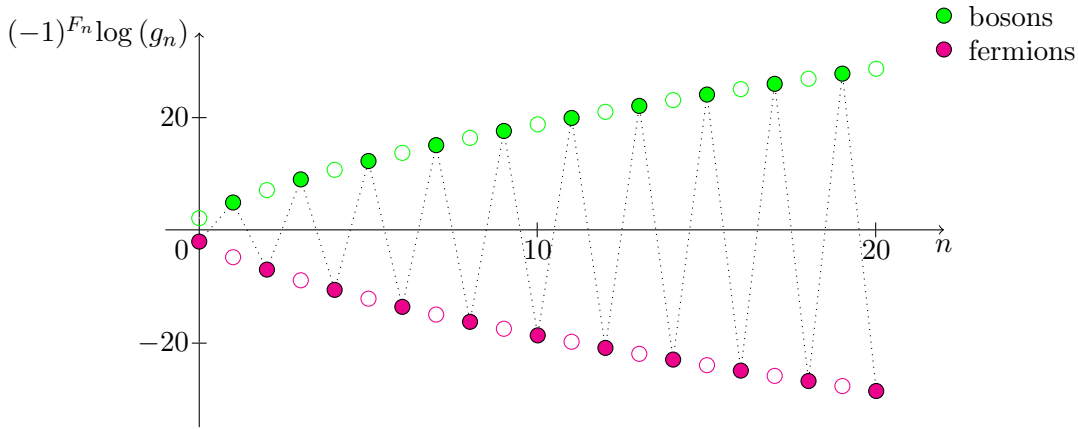
\begin{figure}[ht]
    \centering
    
    \begin{tikzpicture}[xscale=0.45,yscale=0.075,bos/.style={draw,circle,minimum size=2mm,inner sep=0pt,outer sep=0pt,black,fill=green,solid},fer/.style={draw,circle,minimum size=2mm,inner sep=0pt,outer sep=0pt,black,fill=magenta,solid},gbos/.style={draw,circle,minimum size=2mm,inner sep=0pt,outer sep=0pt,green,solid},gfer/.style={draw,circle,minimum size=2mm,inner sep=0pt,outer sep=0pt,magenta,solid},zer/.style={draw,circle,minimum size=2mm,inner sep=0pt,outer sep=0pt,black,fill=orange,solid}]

    \draw[white] (0,0) -- (34,0);
    
    \draw (-1,0) -- (0,0) node[below left]{$0$};
    \draw[-|] (0,0) -- (10,0) node[below]{$10$};
    \draw[-|] (10,0) -- (20,0) node[below]{$20$};
    \draw[->] (20,0) -- (22,0) node[below]{$n$};
    \draw[-|] (0,-35) -- (0,-20) node[left]{$-20$};
    \draw[-|] (0,-20) -- (0,20) node[left]{$20$};
    \draw[->] (0,20) -- (0,35) node[left]{$(-1)^{F_n} \mathrm{log} \, (g_n)$};
    
    \draw [dotted,white] (0,2.07944154) node[gbos]{} -- (2,7.049254841255837) node[gbos]{} -- (4,10.648088014684989) node[gbos]{} -- (6,13.659502153634902) node[gbos]{} -- (8,16.31417809100231) node[gbos]{} -- (10,18.720586679634717) node[gbos]{} -- (12,20.940576396593297) node[gbos]{} -- (14,23.013556682034682) node[gbos]{} -- (16,24.966548083601058) node[gbos]{} -- (18,26.81907090081016) node[gbos]{} -- (20,28.585792100987767) node[gbos]{};
    
    \draw [dotted,white] (1,-4.852030263919617) node[gfer]{} -- (3,-8.946374826141717) node[gfer]{} -- (5,-12.208310140470365) node[gfer]{} -- (7,-15.023099217200523) node[gfer]{} -- (9,-17.543847620246574) node[gfer]{} -- (11,-19.851049252007453) node[gfer]{} -- (13,-21.99353602270788) node[gfer]{} -- (15,-24.003693365249816) node[gfer]{} -- (17,-25.90435950977719) node[gfer]{} -- (19,-27.71238242905967) node[gfer]{};

    \draw [dotted] (0,-2.07944154) node[fer]{} -- (1,4.852030263919617) node[bos]{} -- (2,-7.049254841255837) node[fer]{} -- (3,8.946374826141717) node[bos]{} -- (4,-10.648088014684989) node[fer]{} -- (5,12.208310140470365) node[bos]{} -- (6,-13.659502153634902) node[fer]{} -- (7,15.023099217200523) node[bos]{} -- (8,-16.31417809100231) node[fer]{} -- (9,17.543847620246574) node[bos]{} -- (10,-18.720586679634717) node[fer]{} -- (11,19.851049252007453) node[bos]{} -- (12,-20.940576396593297) node[fer]{} -- (13,21.99353602270788) node[bos]{} -- (14,-23.013556682034682) node[fer]{} -- (15,24.003693365249816) node[bos]{} -- (16,-24.966548083601058) node[fer]{} -- (17,25.90435950977719) node[bos]{} -- (18,-26.81907090081016) node[fer]{} -- (19,27.71238242905967) node[bos]{} -- (20,-28.585792100987767) node[fer]{};
    
    \node[bos] at (22,38){};
    \node[right] at (22,38){\, bosons};
    \node[fer] at (22,32){};
    \node[right] at (22,32){\, fermions};

    \end{tikzpicture}
    
    \caption{The net number of bosonic and fermionic physical degrees of freedom for the lightest energy levels for an anti-D$p$-brane on top of an O$p$-plane, defined as $(-1)^{F_n} g_n = N_b(n) - N_f(n)$, with $N_b(n)$ and $N_f(n)$ being the number of bosonic and fermionic states at the $n$-th mass level. Each point corresponds to states with mass $M_n^2 = n / \alpha'$, with $n=0,1,\dots,20$. Filled points correspond to states that are invariant under the orientifold projection, whereas empty dots represent states that would be there if the anti-D$p$-brane was at a smooth point, with a supersymmetric matching in the bosonic and fermionic degrees of freedom.}
    
    \label{anti-D/O plot}
    
\end{figure}

Misaligned supersymmetry \cite{Dienes:1994np, Dienes:1995pm, Dienes:2001se} is an idea describing string-theory models that are not supersymmetric, and therefore lack a one-to-one matching between bosonic and fermionic number of states at each energy level. Instead, these theories have an exponentially growing oscillation between the net number of bosons and fermions at each mass level.  A simple example realising this property is made up by an anti-D$p$-brane on top of an O$p$-plane \cite{Cribiori:2020sct}, whose spectrum is sketched in fig.~\ref{anti-D/O plot}. The proposal of refs.~\cite{Dienes:1994np, Dienes:1995pm, Dienes:2001se} is that misaligned theories nonetheless have observables that undergo boson-fermion cancellations because all of the infinitely-many contributions average out.

In order to assess whether a misaligned theory actually exhibits boson-fermion cancellations that keep physical observables finite (at least at one-loop), one needs control over the number of degrees of freedom at each mass level. In string theory, such a number is counted by the coefficients appearing in the $q$-expansion of the partition function, where $q=\e^{2 \pi \I \tau}$, with $\tau = \tau_1 + \I \tau_2$ being the modular parameter of the theory. As discussed by refs.~\cite{sussman2017rademacher, Cribiori:2020sct}, these coefficients are particularly simple to determine if the partition function can be expressed as a quotient of Dedekind $\eta$-functions, since they can be obtained via a simple Hardy-Ramanujan-Rademacher expansion. After a review of such expansion in subsection \ref{subsec: sussman}, in subsection \ref{subsec: envelope functions} we review the idea of sector-average, which is the tool commonly used in the literature to discuss the presence of misaligned supersymmetry.

\subsection[Hardy-Ramanujan-Rademacher expansions for simple Dedekind eta-quotients]{Hardy-Ramanujan-Rademacher expansions for simple Dedekind \texorpdfstring{$\boldsymbol{\eta}$}{$\eta$}-quotients} \label{subsec: sussman}
In this subsection, we summarize the Hardy-Ramanujan-Rademacher expansions for the net boson-fermion state degeneracies in partition functions composed of a class of Dedekind $\eta$-quotients, as discussed in ref.~\cite{sussman2017rademacher}. Let $\lbrace \delta_m \rbrace_{m=1}^\infty$ be a sequence of integers $\delta_m \in \mathbb{Z}$ with only finitely many non-vanishing ones. Let $Z=Z(\tau)$ then be the Dedekind $\eta$-quotient 
\begin{equation} \label{Z}
    Z(\tau) = \prod_{m=1}^\infty \bigl[ \eta (m \tau) \bigr]^{\delta_m} = q^{-n_0} \sum_{n=0}^\infty a_n q^n,
\end{equation}
where $a_n$ represent the Laurent coefficients in the expansion in terms of the variable $q = \e^{2 \pi \I \tau}$. 
Let the constants $n_0$, $c_1$ and the functions $c_2=c_2(\alpha)$, $c_3=c_3(\alpha)$ be defined as
\begin{subequations} \label{E:qcoeffs}
\begin{align}
    n_0 & = - \dfrac{1}{24} \sum_{m=1}^\infty m \, \delta_m,\\
    c_1 & = - \dfrac{1}{2} \sum_{m=1}^\infty \delta_m, \label{c1} \\
    c_2(\alpha) & = \prod_{m=1}^\infty \biggl[ \dfrac{\mathrm{gcd} (m,\alpha)}{m} \biggr]^{\frac{\delta_m}{2}}, \\
    c_3(\alpha) & = - \sum_{m=1}^\infty \delta_m \, \dfrac{[\mathrm{gcd} (m,\alpha)]^2}{m}.
\end{align}
\end{subequations}
Then, given the Dedekind sum
\begin{equation}
    s(k,\alpha) = \sum_{n=1}^{\alpha-1} \dfrac{n}{\alpha} \biggl( \dfrac{k n}{\alpha} - \biggl\lfloor \dfrac{k n}{\alpha} \biggr\rfloor - \dfrac{1}{2} \biggr)
\end{equation}
and the function
\begin{equation}
    \varphi(k,\alpha) = \e^{-\I \pi \, \sum_{m=1}^\infty \delta_m \, s \left( \frac{m k}{\mathrm{gcd} \, (m,\alpha)}, \frac{\alpha}{\mathrm{gcd} \, (m,\alpha)} \right)},
\end{equation}
let the function $P_\alpha=P_\alpha(n)$ be
\begin{equation} \label{A_k(n)}
    P_\alpha(n) = \sum_{\substack{0 \leq k < \alpha, \\ \mathrm{gcd} \, (k,\alpha) = 1}} \e^{- 2 \pi \I n \frac{k}{\alpha}} ~ \varphi(k,\alpha).
\end{equation}
Finally, let the function $G=G(\alpha)$ be
\begin{equation}
\label{G(k)}
    G(\alpha) = \underset{ m \in \mathbb{N}: \; \delta_m \neq 0 }{\mathrm{min}} \,\biggl\lbrace \dfrac{[\mathrm{gcd} \, (m,\alpha)]^2}{m} \biggr\rbrace - \dfrac{c_3(\alpha)}{24}.
\end{equation}
With these definitions in hand, the main result of ref.~\cite{sussman2017rademacher} is the following theorem.

\emph{Theorem.} If $c_1 > 0$ and $G(\alpha)$ is a non-negative function, then, for an arbitrary positive integer $n$ that satisfies $n > n_0$, the coefficients $a_n$ in the series expansion of the function $Z(\tau)$ can be written as
\begin{equation} \label{d(n)}
    a_n = \sum_{\substack{\alpha \in \mathbb{N}, \\ c_3(\alpha) > 0}} \dfrac{2\pi \, c_2(\alpha) \, [c_3(\alpha)]^{\frac{c_1+1}{2}}}{[24(n-n_0)]^{\frac{c_1+1}{2}}} \, \dfrac{P_\alpha(n)}{\alpha} \, I_{c_1+1} \biggl[ \biggl( \dfrac{2 \pi^2}{3 \alpha^2} \, c_3(\alpha) (n-n_0) \biggr)^{\frac{1}{2}} \biggr],
\end{equation}
where $I_{\delta}(z)$ is the modified Bessel function of the first kind.

The theorem of eq.~\eqref{d(n)} can be used to compute exactly the net degeneracies for physically interesting partition functions. For example, for an anti-D$p$-brane on an O$p$-plane, after a few manipulations detailed in ref.~\cite{Cribiori:2020sct}, one arrives at the exponentially growing oscillations pictured in fig. \ref{anti-D/O plot}. Note that, because of the asymptotic expansion $\smash{I_\nu(x) \overset{x \sim \infty}{\simeq} \e^x/(2 \pi x)^{\frac{1}{2}}}$, one observes in eq.~\eqref{d(n)} the leading Hagedorn behaviour and, moreover, each decreasing value $c_3(\alpha)/\alpha^2$ represents a successively subleading exponential correction to the coefficient $a_n$.

The series coefficients $a_n$ in eq.~\eqref{d(n)} involve the $n$-dependent, periodic functions $P_\alpha(n)$. For a fixed $\alpha$, the $P_\alpha(n)$ can only take up to $\alpha$ different values, which we denote as $P_{\alpha}(\beta)$, with $\beta=1,\dots,\alpha$. One can prove the following lemma \cite{Cribiori:2020sct}.

\emph{Lemma.} 
Given the integers $m$, $\alpha\in \mathbb{N}$, $n\in \mathbb{N}_0$ and for $\gamma = \mathrm{gcd}  (\alpha,m)$, if $\nexists \, p \in \mathbb{N}: \; m = p\,\alpha$, i.e., if $m$ is not a multiple of $\alpha$ and if $\alpha>1$, then
\begin{equation} \label{closure}
    \sum_{\beta=0}^{\frac{\alpha}{\gamma}-1} P_\alpha(n+m\beta) = 0.
\end{equation}
This lemma has been used in ref.~\cite{Cribiori:2020sct} to prove the presence of misaligned supersymmetry at all orders in the Hardy-Ramanujan-Rademacher expansion, as we will now review.

\subsection[The sector-averaged <an>]{The sector-averaged \texorpdfstring{$\boldsymbol{\langle a_n \rangle}$}{$\langle a_n \rangle$}} \label{subsec: envelope functions}
\begin{figure}[ht]
    \centering
    
    \begin{tikzpicture}[xscale=0.45,yscale=0.075,bos/.style={draw,circle,minimum size=2mm,inner sep=0pt,outer sep=0pt,black,fill=green,solid},fer/.style={draw,circle,minimum size=2mm,inner sep=0pt,outer sep=0pt,black,fill=magenta,solid},gbos/.style={draw,circle,minimum size=2mm,inner sep=0pt,outer sep=0pt,thick,green,solid},gfer/.style={draw,circle,minimum size=2mm,inner sep=0pt,outer sep=0pt,thick,magenta,solid},zer/.style={draw,circle,minimum size=2mm,inner sep=0pt,outer sep=0pt,black,fill=orange,solid},corr1/.style={draw,circle,minimum size=1.5mm,inner sep=0pt,outer sep=0pt,black,fill=yellow,solid},corr2/.style={draw,circle,minimum size=1.5mm,inner sep=0pt,outer sep=0pt,black,fill=orange,solid},corr3/.style={draw,circle,minimum size=1.5mm,inner sep=0pt,outer sep=0pt,black,fill=cyan,solid}]
    
    \draw[domain=2:22.5, smooth, thick, variable=\n, gray] plot ({\n}, {- ln(8*8^(1/4)) - (11/4)*ln(\n) + (8*pi^2)^(1/2)*\n^(1/2)}) node[right, black]{$\Phi_1(n)$};
    \draw[domain=2:22.5, smooth, thick, variable=\n, gray] plot ({\n}, {ln(8*8^(1/4)) + (11/4)*ln(\n) - (8*pi^2)^(1/2)*\n^(1/2)}) node[right, black]{$-\Phi_1(n)$};
    
    \draw[domain=1.75:22.5, dashed, variable=\n, gray] plot ({\n}, {- ln(8*8^(1/4)) - (11/4)*ln(\n) + ((8*pi^2)^(1/2)-1.5)*\n^(1/2)}) node[below, black, xshift=5.0, yshift=-2.5]{$\Phi_{3}(n;1)=\Phi_{3}(n;3)$};
    \draw[domain=1.75:19.0, dashed, variable=\n, gray] plot ({\n}, {- ln(8*8^(1/4)) - (11/4)*ln(\n) + ((8*pi^2)^(1/2)+3)*\n^(1/2)}) node[below, black, xshift=5.0, yshift=-2.0]{$\Phi_{3}(n;2)$};
    
    \draw[white] (0,0) -- (34,0);
    \draw[white] (0,0) -- (0,44);
    
    \draw (-1,0) -- (0,0) node[below left]{$0$};
    \draw[-|] (0,0) -- (10,0) node[below]{$10$};
    \draw[-|] (10,0) -- (20,0) node[below]{$20$};
    \draw[->] (20,0) -- (22,0) node[below]{$n$};
    \draw[-|] (0,-35) -- (0,-20) node[left]{$-20$};
    \draw[-|] (0,-20) -- (0,20) node[left]{$20$};
    \draw[->] (0,20) -- (0,35) node[left]{$(-1)^{F_n} \mathrm{log} \, (g_n)$};
    
    \draw [dotted] (0,-2.07944154) node[fer]{} -- (1,4.852030263919617) node[bos]{} -- (2,-7.049254841255837) node[fer]{} -- (3,8.946374826141717) node[bos]{} -- (4,-10.648088014684989) node[fer]{} -- (5,12.208310140470365) node[bos]{} -- (6,-13.659502153634902) node[fer]{} -- (7,15.023099217200523) node[bos]{} -- (8,-16.31417809100231) node[fer]{} -- (9,17.543847620246574) node[bos]{} -- (10,-18.720586679634717) node[fer]{} -- (11,19.851049252007453) node[bos]{} -- (12,-20.940576396593297) node[fer]{} -- (13,21.99353602270788) node[bos]{} -- (14,-23.013556682034682) node[fer]{} -- (15,24.003693365249816) node[bos]{} -- (16,-24.966548083601058) node[fer]{} -- (17,25.90435950977719) node[bos]{} -- (18,-26.81907090081016) node[fer]{} -- (19,27.71238242905967) node[bos]{} -- (20,-28.585792100987767) node[fer]{};
    
    \draw[dotted, red] (1,4.852030263919617-1.5) node[corr1]{} -- (3,8.946374826141717-1.5*1.73) node[corr2]{} -- (5,12.208310140470365+3*2.24) node[corr3]{} -- (7,15.023099217200523-1.5*2.64) node[corr1]{} -- (9,17.543847620246574-1.5*3) node[corr2]{} -- (11,19.851049252007453+3*3.32) node[corr3]{} -- (13,21.99353602270788-1.5*3.61) node[corr1]{} -- (15,24.003693365249816-1.5*3.87) node[corr2]{} -- (17,25.90435950977719+3*4.12) node[corr3]{} -- (19,27.71238242905967-1.5*4.36) node[corr1]{};
    
    \draw[dotted, red] (2,-7.049254841255837-3*1.41) node[corr3]{} -- (4,-10.648088014684989+1.5*2) node[corr1]{} -- (6,-13.659502153634902+1.5*2.45) node[corr2]{} -- (8,-16.31417809100231-3*2.83) node[corr3]{} -- (10,-18.720586679634717+1.5*3.16) node[corr1]{} -- (12,-20.940576396593297+1.5*3.46) node[corr2]{} -- (14,-23.013556682034682-3*3.74) node[corr3]{} -- (16,-24.966548083601058+1.5*4) node[corr1]{} -- (18,-26.81907090081016+1.5*4.24){} node[corr2]{} -- (20,-28.585792100987767-3*4.47) node[corr3]{};
    
    \node[corr1] at (21.5,58){};
    \node[right] at (21.5,58){\, $P_3(1)=-1$};
    \node[corr3] at (21.5,51){};
    \node[right] at (21.5,51){\, $P_3(2)=+2$};
    \node[corr2] at (21.5,44){};
    \node[right] at (21.5,44){\, $P_3(3)=-1$};

    \end{tikzpicture}
    
    \caption{A schematic plot representing the spectrum of an anti-D$p$-brane on top of an O$p$-plane, including the terms at leading order, for $\alpha=1$, and the (magnified) corrections at next-to-leading order, for $\alpha=3$. One has to consider bosons (odd $n$) and fermions (even $n$) separately. Then, levels $n = 1 \, \mathrm{mod} \, 3$ have corrections multiplied by the value $\smash{P_3(1)=-1}$, levels $n = 2 \, \mathrm{mod} \, 3$ have corrections multiplied by the value $\smash{P_3(2)=+2}$ and levels $n = 3 \, \mathrm{mod} \, 3$ have corrections multiplied by the value $\smash{P_3(3)=-1}$. For each different value the function $\smash{P_\alpha(n)}$ can take, one can individuate a different interpolating function, both for bosons and for fermions. Evidently, the average of such interpolating functions vanishes, in both the bosonic sector and in the fermionic sector, separately.}
    
    \label{corrections to anti-D/O plot}
    
\end{figure}
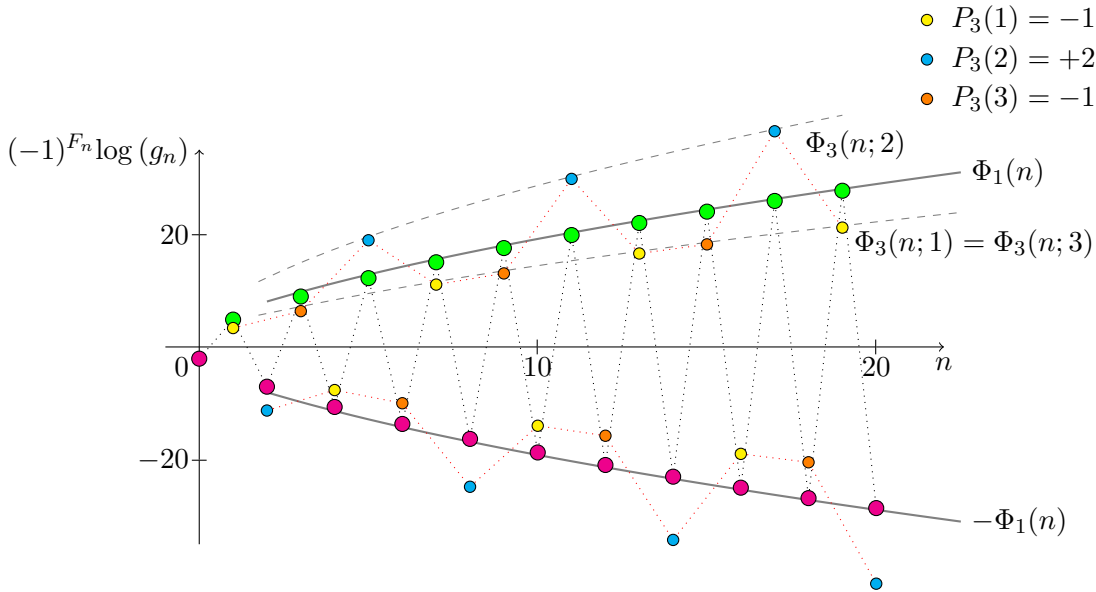

In misaligned string theories, one can use the Hardy-Ramanujan-Rademacher expansion in eq.~(\ref{d(n)}) to compute the physical state net-degeneracies in different sectors, $a^i_n$, corresponding to distinct discrete sets for $n$ (for instance, in fig. \ref{anti-D/O plot} for an anti-D$p$-brane on an O$p$-plane, one observes two sectors, $n$ even and $n$ odd corresponding to fermionic and bosonic abundances, respectively). Each sector's net-degeneracy grows as $a^i_n \sim A \, n^{-B} e^{C \sqrt{n}}$ for large $n$, with $C$ the inverse Hagedorn temperature and $A$ and $B$ constants.  However, by analytically continuing the net degeneracies $a^i_n$ to continuous $n$, introducing the envelope functions $\Phi_i(n)$ for $n \in \mathbb{R}$, a modular-invariant sector-averaged net-degeneracy, $\langle a_n \rangle \equiv \sum_i \Phi_i(n)$, can be defined, in which the exponentially growing net boson-fermion oscillations lead to cancellations \cite{Dienes:1994np}. Indeed, ref.~\cite{Dienes:1994np} showed that for oriented closed-string theories that are modular-invariant and tachyon-free, the $\alpha=1$ leading order exponential growth in the Hardy-Ramanujan-Rademacher expansion cancels in the sector-average, leaving a slower growth $\langle a_n \rangle  \sim A \, n^{-B} e^{C_{\mathrm{eff}} \sqrt{n}}$ for large $n$, with $C_{\mathrm{eff}} < C$. It was moreover conjectured that $C_{\mathrm{eff}}=0$. In ref. \cite{Cribiori:2020sct}, we extended these results to open-string models. Moreover, by extending the notion of sector-average to include an average over subleading contributions at each order $\alpha>1$ in the Hardy-Ramanujan-Rademacher expansion, and using lemma \eqref{closure}, we proved that $C_{\mathrm{eff}}=0$. This is illustrated in fig. \ref{corrections to anti-D/O plot} for the case of an anti-D$p$-brane on an O$p$-plane.

Although these cancellations associated with misaligned supersymmetry seem remarkable, the physical significance of the envelope functions and sector-averages introduced in refs.~\cite{Dienes:1994np, Cribiori:2020sct} was not clear. The purpose of this paper is to demonstrate that the same cancellations occur in physically meaningful quantities, like the one-loop vacuum energy.

\subsection{Misaligned supersymmetry and modular transformations}
A fundamental result of ref.~\cite{Dienes:1994np} is the proof that, for closed strings, a sufficient condition for the envelope functions to average out to zero, at leading order in the Hardy-Ramanujan-Rademacher sum, is the modular invariance of the partition function $Z=Z(\tau,\overline{\tau})$. This conclusion holds whenever the theory with a modular-invariant partition function is non-supersymmetric and free of physical tachyons.

In ref.~\cite{Cribiori:2020sct}, for a class of theories, the cancellation of the envelope functions in the sector-average has been shown at all orders in the Hardy-Ramanujan-Rademacher sum. This applies to both closed and open strings.  
Whilst the role of modular invariance in closed strings at one-loop is clear,\footnote{In fact, we will see below that misaligned cancellations for closed strings take place piece-wise within modular non-invariant terms, similar to the open-string cancellations, though in the closed-string case the non-invariant terms add up to make a fully modular partition function.} its appearance for one-loop open-string diagrams may be puzzling. It is therefore worthwhile to spend a few words to recall how open-string partition functions are covariant under some subgroup of the full modular group, and moreover, how the well-known open-closed string duality and electric-magnetic D$p$/D$(6-p)$ duality can be expressed in terms of modular transformations.

\subsubsection*{Open strings and modular symmetry}

Open-string models descend from left-right symmetric closed-string models after a worldsheet (parity) orbifolding that mixes left and right movers, together with a further target-space $\mathbb{Z}_2$-involution for orientifolds (see ref.~\cite{Angelantonj:2002ct} for a review). In addition to these open-string descendants, further open sectors can be introduced via probe D-branes. A consequence of the orbifolding is that the modular invariance of the closed string at one-loop is broken. However, as we will now discuss, a remnant symmetry survives.

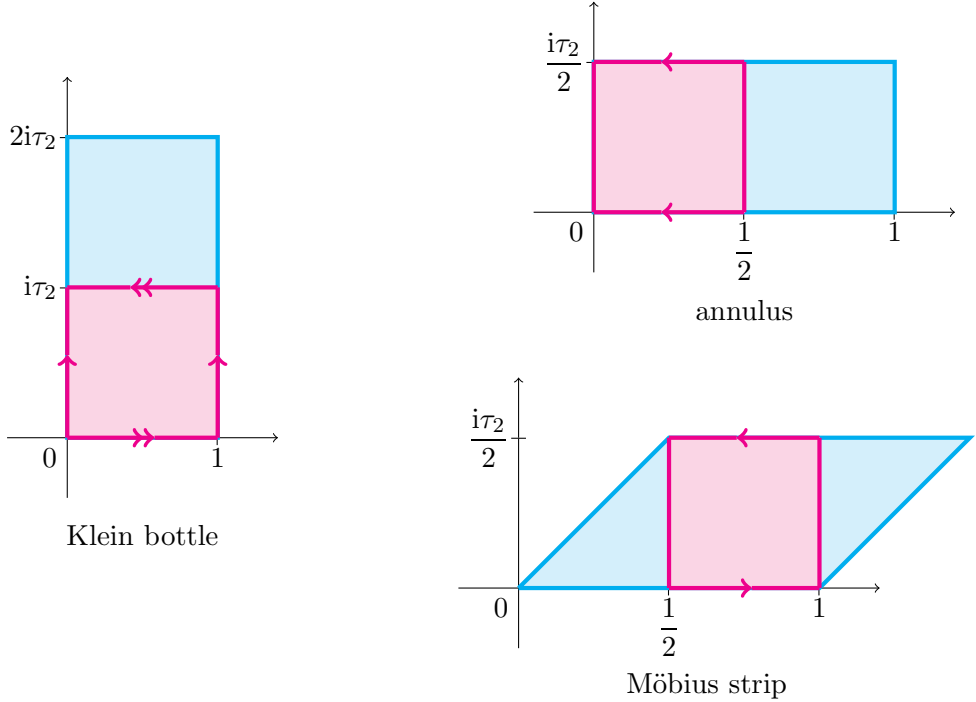
\begin{figure}[ht]
    \centering
    
    \begin{tikzpicture}[xscale=2,yscale=2]
    
    \def \xA {3.5}
    \def \yA {1.5}
    
    \def \xM {3}
    \def \yM {-1}
    
    \draw (-0.4,0) -- (0,0) node[below left]{$0$};
    \draw[-|] (0,0) -- (1,0) node[below]{$1$};
    \draw[->] (1,0) -- (1.4,0);
    \draw (0,-0.4) -- (0,0);
    \draw[-|] (0,0) -- (0,1) node[left]{$\I \tau_2$};
    \draw[-|] (0,1) -- (0,2) node[left]{$2\I \tau_2$};
    \draw[->] (0,2) -- (0,2.4);

    
    \draw[ultra thick, color=cyan, fill=cyan!15!white] (0,0) rectangle (1,2);
    
    \fill[magenta!20!white] (0,0) rectangle (1,1);
    \draw[->>,ultra thick,color=magenta] (0,0) -- (0.58,0);
    \draw[ultra thick,color=magenta] (0.58,0) -- (1,0);
    \draw[->,ultra thick,color=magenta] (1,0) -- (1,0.55);
    \draw[ultra thick,color=magenta] (1,0.55) -- (1,1);
    \draw[->>,ultra thick,color=magenta] (1,1) -- (0.42,1);
    \draw[ultra thick,color=magenta] (0.42,1) -- (0,1);
    \draw[->,ultra thick,color=magenta] (0,0) -- (0,0.55);
    \draw[ultra thick,color=magenta] (0,0.55) -- (0,1);
    
    \node at (0.5,-0.65){Klein bottle};

    
    \draw (-0.4+\xA,0+\yA) -- (0+\xA,0+\yA) node[below left]{$0$};
    \draw[-|] (0+\xA,0+\yA) -- (1+\xA,0+\yA) node[below]{$\dfrac{1}{2}$};
    \draw[-|] (1+\xA,0+\yA) -- (2+\xA,0+\yA) node[below]{$1$};
    \draw[->] (2+\xA,0+\yA) -- (2.4+\xA,0+\yA);
    \draw (0+\xA,-0.4+\yA) -- (0+\xA,0+\yA);
    \draw[-|] (0+\xA,0+\yA) -- (0+\xA,1+\yA) node[left]{$\dfrac{\I \tau_2}{2}$};
    \draw[->] (0+\xA,1+\yA) -- (0+\xA,1.4+\yA);
    
    \draw[ultra thick, color=cyan, fill=cyan!15!white] (0+\xA,0+\yA) rectangle (2+\xA,1+\yA);
    
    \fill[magenta!20!white] (0+\xA,0+\yA) rectangle (1+\xA,1+\yA);
    \draw[->,ultra thick,color=magenta] (1+\xA,0+\yA) -- (0.45+\xA,0+\yA);
    \draw[ultra thick,color=magenta] (0.45+\xA,0+\yA) -- (0+\xA,0+\yA);
    \draw[ultra thick,color=magenta] (1+\xA,0+\yA) -- (1+\xA,1+\yA);
    \draw[->,ultra thick,color=magenta] (1+\xA,1+\yA) -- (0.45+\xA,1+\yA);
    \draw[ultra thick,color=magenta] (0.45+\xA,1+\yA) -- (0+\xA,1+\yA);
    \draw[ultra thick,color=magenta] (0+\xA,0+\yA) -- (0+\xA,1+\yA);
    
    \node at (1+\xA,-0.65+\yA){annulus};
    
    
    \draw (-0.4+\xM,0+\yM) -- (0+\xM,0+\yM) node[below left]{$0$};
    \draw[-|] (0+\xM,0+\yM) -- (1+\xM,0+\yM) node[below]{$\dfrac{1}{2}$};
    \draw[-|] (1+\xM,0+\yM) -- (2+\xM,0+\yM) node[below]{$1$};
    \draw[->] (2+\xM,0+\yM) -- (2.4+\xM,0+\yM);
    \draw (0+\xM,-0.4+\yM) -- (0+\xM,0+\yM);
    \draw[-|] (0+\xM,0+\yM) -- (0+\xM,1+\yM) node[left]{$\dfrac{\I \tau_2}{2}$};
    \draw[->] (0+\xM,1+\yM) -- (0+\xM,1.4+\yM);
    
    \draw[ultra thick, color=cyan, fill=cyan!15!white] (0+\xM,0+\yM) -- (2+\xM,0+\yM) -- (3+\xM,1+\yM) -- (1+\xM,1+\yM) -- (0+\xM,0+\yM);
    
    \fill[magenta!20!white] (1+\xM,0+\yM) rectangle (2+\xM,1+\yM);
    \draw[->,ultra thick,color=magenta] (1+\xM,0+\yM) -- (1.55+\xM,0+\yM);
    \draw[ultra thick,color=magenta] (1.55+\xM,0+\yM) -- (2+\xM,0+\yM);
    \draw[ultra thick,color=magenta] (2+\xM,0+\yM) -- (2+\xM,1+\yM);
    \draw[->,ultra thick,color=magenta] (2+\xM,1+\yM) -- (1.45+\xM,1+\yM);
    \draw[ultra thick,color=magenta] (1.45+\xM,1+\yM) -- (1+\xM,1+\yM);
    \draw[ultra thick,color=magenta] (1+\xM,0+\yM) -- (1+\xM,1+\yM);
    
    \node at (1.25+\xM,-0.65+\yM){Möbius strip};

    \end{tikzpicture}
    
    \caption{A sketch of the fundamental polygons (magenta) and of the double-covering tori (cyan) for the Klein bottle, the annulus and the Möbius strip in the complex plane. Any point $z \in \mathbb{C}$ can be mapped into the fundamental polygon by means of a combination of the corresponding antiholomorphic involution and of the lattice symmetry of the associated double-covering torus.}
    
    \label{fundamental polygons + double-covering tori}
    
\end{figure}

For open-string models, which of course include also closed-string sectors, four worldsheet surfaces contribute to the one-loop vacuum amplitude: the torus (closed orientable), the Klein bottle (closed non-orientable), the annulus (open orientable) and the Möbius strip (open non-orientable). The latter three surfaces can each be described in terms of closed orientable double-covering tori \cite{Alessandrini:1971cz, Alessandrini:1971dd} (see also e.g.~refs.\cite{Antoniadis:1996vw, Angelantonj:2002ct}), with the complex structures parametrised as usual by a value $\tau \in \mathbb{C}$. The original fundamental polygons $P$ are then recovered by quotienting the double-covering tori under associated anti-conformal involutions $I(z)$. A convenient set of parametrizations is as follows \cite{Bianchi:1988ux, Sagnotti:1987tw, Antoniadis:1996vw}:
\begin{subequations}
    \begin{empheq}[]{align}
    & P_{\mathrm{K}} = [0,1] \times [0, \I \tau_2]: && I_{\mathrm{K}}(z) = 1 - \overline{z} + \I \tau_2, && \tau = 2 \I \tau_2; \\[0.5ex]
    & P_{\mathrm{A}} = \Bigl[0,\dfrac{1}{2}\Bigr] \times \Bigl[0, \dfrac{\I \tau_2}{2}\Bigr]: && I_{\mathrm{A}}(z) = - \overline{z} = 1 - \overline{z}, && \tau=\dfrac{\I\tau_2}{2}; \\
    & P_{\mathrm{M}} = \Bigl[\dfrac{1}{2},1\Bigr] \times \Bigl[0, \dfrac{\I \tau_2}{2}\Bigr]: && I_{\mathrm{M}}(z) = \dfrac{1}{2} - \overline{z} + \dfrac{\I \tau_2}{2} && \tau = \dfrac{\I \tau_2}{2} + \dfrac{1}{2}.
    \end{empheq}
\end{subequations}
For a sketch of the fundamental polygons together with their double-covering tori, see fig.~\ref{fundamental polygons + double-covering tori}. 

In particular, note that in order to obtain the fundamental polygons from the double-covering tori via the involutions, the complex structures of the double-tori are fixed to be $2\I \tau_2$ for the Klein bottle, $\I \tau_2/2$ for the annulus, and $1/2 + \I \tau_2/2$ for the Möbius strip.  Each double-torus thus has only one modular parameter and its surviving modular group is trivial. The Klein-bottle, annulus and M\"obius-strip amplitudes are naturally expressed in terms of these moduli of the respective double-tori, with the domain of integration extending along the whole positive imaginary axis of the $\tau$-plane (see e.g.~refs.~\cite{CasteloFerreira:2000mz, Antoniadis:2005sd}). It is interesting to note that these observations have been generalised to higher genus-$g$ surfaces.  For $g >1$, the modular transformations that preserve the involution form a non-trivial subgroup of $\mathrm{Sp}(2g,\mathbb{Z})$, called the ``relative modular group'' \cite{Bianchi:1989du}.

Although the orbifolding to reach the descendant genus-one surfaces breaks the modular invariance of the covering closed oriented tori, modular transformations still have some role to play. Indeed, a modular transformation underlies the famous open-closed string duality. Let us focus on the M\"obius strip, as this will be the main diagram of interest in what follows. The M\"obius strip can itself be interpreted as a dual tree-level closed-string diagram, with the different channels being related by the modular $\mathrm{P}$-transformation \cite{Pradisi:1988xd, Angelantonj:2002ct}, with $P=T S T^2 S$. Moreover, the interaction between a probe D$p$-brane and an O$p$-plane, given by the Möbius-strip amplitude, is related to a D$(6-p)$-brane/O$(6-p)$-plane interaction by an $\mathrm{S}$-transformation (see e.g.~ref.~\cite{Cribiori:2020sct}). As we will see, in the end our open-string partition functions will carry an covariance under congruence subgroups of $\mathrm{PSL}_2(\mathbb{Z})$.

As an example, the Dedekind $\eta$-quotients set up in eqs.~(\ref{Z}) and (\ref{E:qcoeffs}), are covariant forms of weight $k=-c_1$ under the congruence $\mathrm{PSL}_2(\mathbb{Z})$-subgroup with a generically non-trivial multiplier system \cite{sussman2017rademacher}
\begin{equation}
    \Gamma_0 (n) = \Biggl\lbrace \left( \begin{array}{cc}
        a & b \\
        c & d
    \end{array} \right) \in \mathrm{PSL}_2(\mathbb{Z}): \, c = 0 \, \mathrm{mod} \, n \Biggr\rbrace,
\end{equation}
where $n = \mathrm{lcm} \, \lbrace m \in \mathbb{N}: \, \delta_m \neq 0 \rbrace$.   Therefore, the amplitude for the anti-D$p$-brane/O$p$-plane case, which takes the form $M(\tau) = - 8 \, \eta^{16}(\tau) \eta^{16}(4\tau) / \eta^{40}(2 \tau)$, is covariant, with weight $k=-4$, under the subgroup $\Gamma_0 (4)$.  Note moreover that, although the $\mathrm{S}$-transformation is not part of $\Gamma_0(n)$ for $n\neq1$, the Dedekind quotient $M(\tau) = \prod_m [\eta(m \tau)]^{\delta_m}$ transforms under the $\mathrm{S}$-transformation as $\smash{M(-1/\tau) = (-\I \tau)^{-c_1} (\prod_m m^{-\delta_m/2}) \prod_m [\eta(\tau/m)]^{\delta_m}}$, where we may still interpret $-c_1$ as a `weight'.  In the case of a D$p$-brane on an O$p$-plane, this $\mathrm{S}$-transformation realises the electric-magnetic duality between D$p$- and D$(6-p)$-branes, and will be discussed further around eq. (\ref{anti-Dp/Op S-transformation}).

\section{One-loop cosmological constant in string theory} \label{sec: Lambda}

Since the cosmological constant is the main observable we are interested in, in this section we review its definition in perturbative string theory at one loop and point out the aspects relevant for our analysis.

Let us consider a $D$-dimensional quantum field theory consisting of a tower of string states, labelled by a discrete index $n$, with mass levels $\smash{M^2_n}$. We denote the number of bosonic states minus the number of fermionic states at level $n$ by $(-1)^{F_n} g_n$, with $g_n\geq0$, and we call $\smash{g_n}$ the net state degeneracies and $\smash{F_n}$ the fermion parities. Given an arbitrary mass scale $\mu^2$, in terms of a Schwinger proper-time parameter $t$, the one-loop cosmological constant reads \cite{Dienes:1995pm, Dienes:2001se}
\begin{equation} \label{generalLambda}
    \Lambda = - \dfrac{1}{2} \, \biggl(\!\dfrac{\mu^2}{8 \pi^2}\!\biggr)^{\!D/2} \sum_{n} (-1)^{F_n} g_n \int_0^\infty \dfrac{\de t}{t^{1+D/2}} \, \e^{- 2 \pi M^2_n t / \mu^2}.
\end{equation}
In this expression, the region $t \sim \infty$ leads to divergences only in the presence of tachyons, whereas the region $t \sim 0^+$ is instead generally singular unless cancellations occur due to the structure of the net physical degeneracies.

We discuss this expression for open and closed strings in subsections \ref{subsec: openstringLambda} and \ref{subsec: closedstringLambda} below. We will always refer to D$p$-branes when considering open strings.

\subsection{One-loop cosmological constant for open strings} \label{subsec: openstringLambda}
For open strings, the mass spectrum in both the NS- and R-sectors follows the pattern $M^2_n = n/\alpha' $ for each mass level $n \in \mathbb{N}_0$, so it is convenient to set $\mu=1/\sqrt{\alpha'}$. Moreover, for the field theory of a D$p$-brane one must consider a spacetime of dimension $\smash{D=p+1}$. Therefore, eq.~(\ref{generalLambda}) can be rearranged as
\begin{equation} \label{openstringLambda}
    \Lambda_{\mathrm{D}p} = - \dfrac{1}{2 \pi} \, T_{\mathrm{D}p} \int_0^\infty \dfrac{\de t}{2t} \, M_{\mathrm{D}p} (t),
\end{equation}
where the tension of the D$p$-brane is $T_{\mathrm{D}p} = 2 \pi / l_s^{p+1}$, with the string length being $l_s = 2 \pi \sqrt{\alpha'}$, and where we have singled out the partition function
\begin{equation} \label{openstringM}
    M_{\mathrm{D}p} (t) = \dfrac{1}{(2t)^{\frac{1}{2}(p+1)}} \sum_n (-1)^{F_n} g_n \, \e^{-2 \pi t n}.
\end{equation}

In eq.~(\ref{openstringLambda}), the cosmological constant is UV-divergent unless cancellations occur such that the partition function in eq.~(\ref{openstringM}) approaches the origin $t=0$ at least as a power $t^\epsilon$, with $\epsilon>0$. This is the case for supersymmetric theories, where the partition function is identically zero, due to the level-by-level exact matching in the number of fermions and bosons, i.e. $g_n \equiv 0$ for all $n \in \mathbb{N}_0$. As heuristically discussed in ref.~\cite{Cribiori:2020sct}, a similar mechanism is at work in a wider class of theories, where an overall fermion-boson cancellation takes place amongst different levels. This feature is called misaligned supersymmetry and in the next section we are going to show that it is a sufficient condition to have cancellations in the physical contributions to the cosmological constant.

\subsection{One-loop cosmological constant for closed strings} \label{subsec: closedstringLambda}
For closed strings, as the mass spectrum typically follows the pattern $M^2_n = 4n/\alpha'$ for each mass level $n \in \mathbb{N}_0/2$, which is provided by two identical contributions from the right- and left-moving sectors $m_n^2 = \overline{m}_n^2 = 2n/\alpha'$, it is convenient to set $\mu=2/\sqrt{\alpha'}$. Defining a complex variable $\tau = \tau_1 + \I \tau_2$, with $\tau_2 = t/2$, the right-handside in eq.~(\ref{generalLambda}) can accommodate a further integration $\int_{-1/2}^{1/2} \de \tau_1 = 1$. More generally, any term $a_{mn} \, \e^{2 \pi \I \tau_1 (m-n)} \e^{- 2 \pi \tau_2 (m+n)}$ can be added, with $m \neq n$, leaving the result invariant, since the $\tau_1$-integration trivially means $\int_{-1/2}^{1/2} \de \tau_1 \, \e^{2 \pi \I \tau_1 k} = \delta_{k 0}$. Note that this always works since invariance under $\mathrm{T}$-transformations requires $m - n \in \mathbb{Z}$ in string-theory constructions. In particular, defining the variable $q = \e^{2 \pi \I \tau}$, we can express the cosmological constant as
\begin{equation} \label{closedstringLambdaS}
    \Lambda_D = - \dfrac{1}{8 \pi} \dfrac{1}{\kappa_{D}^2 l_s^2} \int_{\mathbb{S}} \dfrac{\de^2 \tau}{\tau_2^2} \, Z(\tau, \overline{\tau}),
\end{equation}
with the $D$-dimensional gravitational coupling constant being $2 \kappa_D^2 = l_s^{D-2} / 2 \pi$, where the partition function, defined as
\begin{equation}\label{closedstringZ}
    Z(\tau, \overline{\tau}) = \tau_2^{1-D/2} \sum_{m} \sum_{n} a_{m n} \, q^m \overline{q}^n,
\end{equation}
is integrated over the domain $\mathbb{S} = \bigl\lbrace \tau \in \mathbb{C}: \; \mathrm{Re} \, \tau \in [-1/2,1/2] \, \wedge  \, \tau_2 \in [0, +\infty[ \bigr\rbrace$ together with the $\mathrm{PSL}_2(\mathbb{Z})$-invariant measure $\de^2 \tau / \tau_2^2$. Note the identification $a_{nn} \equiv (-1)^{F_n} g_n$. The one-loop cosmological constant in eq.~(\ref{closedstringLambdaS}) is free of IR-divergences in the region $\tau_2 \sim \infty$ if the theory is free of physical tachyons. On the other hand, it is UV-divergent in the region $\tau_2 \sim 0^+$. Thanks to modular invariance, this divergence can be removed by restricting the domain of integration to non-redundant configurations.

Indeed, because $Z = Z(\tau, \overline{\tau})$ represents the one-loop partition function of a closed-string theory, it is invariant under the modular group $\mathrm{PSL}_2(\mathbb{Z})$, and the UV-divergence can be interpreted as a gauge divergence. In fact, a manifestly finite result can be obtained by factorising out the redundant volume, restricting the integration to the fundamental domain
\begin{equation}
    \mathbb{F} = \bigl\lbrace \tau \in \mathbb{C}: \;  \tau_1 \in [-1/2,1/2] \, \wedge  \tau_2 \in [0, +\infty[ \, \wedge \, \ab \tau \ab \in [1, +\infty[ \bigr\rbrace.
\end{equation}
Explicitly, therefore, the regularised version of the cosmological constant (\ref{closedstringLambdaS}) reads
\begin{equation} \label{closedstringLambdaF}
    \tilde{\Lambda}_D = - \dfrac{1}{8 \pi} \dfrac{1}{\kappa_{D}^2 l_s^2} \int_{\mathbb{F}} \dfrac{\de^2 \tau}{ \tau_2^2} \, Z(\tau, \overline{\tau}).
\end{equation}
This is an integral definition. Because the singular region corresponding to $\tau_2 = 0$ has been removed, the UV-divergence is absent. In the absence of physical tachyons, this one-loop cosmological constant is finite.

One may also express the regularised one-loop cosmological constant in a different way, by means of the so-called Kutasov-Seiberg identity \cite{Kutasov:1990sv}. Before stating it, we will review a heuristic argument to motivate it. One can account for the gauge divergence in the volume of integration by defining a regulated domain $\mathbb{S}_\sigma = \bigl\lbrace \tau \in \mathbb{C}: \; \mathrm{Re} \, \tau \in [-1/2,1/2] \, \wedge \, \mathrm{Im} \, \tau \in [\sigma^{-1}, +\infty[ \bigr\rbrace$, with $\sigma \gg 1$, and establishing the relationship
\begin{equation} \label{KS-derivation}
    \dfrac{1}{\mathrm{vol}_{\mathrm{PSL}_2(\mathbb{Z})} \, \mathbb{F}} \int_{\mathbb{F}} \dfrac{\de^2 \tau}{\tau_2^2} \; Z(\tau, \overline{\tau}) \overset{\sigma \sim \infty}{\simeq} \dfrac{1}{\mathrm{vol}_{\mathrm{PSL}_2(\mathbb{Z})} \, \mathbb{S}_\sigma} \int_{\mathbb{S}_\sigma} \dfrac{\de^2 \tau}{\tau_2^2} \; Z(\tau, \overline{\tau}),
\end{equation}
where the volumes of $\mathbb{S}_\sigma$ and $\mathbb{F}$ with respect to the modular-invariant measure are
\begin{subequations}
\begin{align}
    \mathrm{vol}_{\mathrm{PSL}_2(\mathbb{Z})} \, \mathbb{S}_\sigma & \equiv \int_{\mathbb{S}_\sigma} \dfrac{\de^2 \tau}{\tau_2^2} = \int_{-1/2}^{1/2} \de \tau_1\int_{\sigma^{-1}}^\infty \dfrac{\de \tau_2}{\tau_2^2} = \sigma, \\
    \mathrm{vol}_{\mathrm{PSL}_2(\mathbb{Z})} \, \mathbb{F} & \equiv \int_{\mathbb{F}} \dfrac{\de^2 \tau}{ \tau_2^2}= \int_{\sqrt{3}/2}^\infty \dfrac{\de \tau_2}{\tau_2^2} - 2 \int_{\sqrt{3}/2}^1 \dfrac{\de \tau_2}{\tau_2^2} \sqrt{1 - \tau_2^2} = \dfrac{\pi}{3}.
\end{align}
\end{subequations}
In the $\mathbb{S}_{\sigma}$-integration, the partition function effectively receives contributions only from the physical states. Defining the function $g(\tau_2)$, which depends only on the net-degeneracies of physical states,
\begin{equation} \label{g(tau2)}
    g(\tau_2) = \int_{-1/2}^{1/2} \de \tau_1 \, Z(\tau_1,\tau_2) = \tau_2^{1-D/2} \sum_{n} a_{nn} \,e^{-4\pi\tau_2 n},
\end{equation}
one can write
\begin{equation}
    \lim_{\sigma \to \infty} \biggl[ \dfrac{1}{\mathrm{vol}_{\mathrm{PSL}_2(\mathbb{Z})} \, \mathbb{S}_\sigma} \int_{\mathbb{S}_\sigma} \dfrac{\de^2 \tau}{\tau_2^2} \, Z (\tau, \overline{\tau}) \biggr] = \lim_{\sigma \to \infty} \dfrac{1}{\sigma} \int_{\sigma^{-1}}^{\infty} \dfrac{\de \tau_2}{\tau_2^2} \, g(\tau_2) = \lim_{\sigma \to \infty} g(\sigma^{-1}),
\end{equation}
assuming the integral to be dominated by the region around $\smash{\tau_2 \sim \sigma^{-1} \sim 0^+}$ and ignoring the $\tau_2$-dependence of $g(\tau_2)$. Putting these expressions together, one arrives at the Kutasov-Seiberg identity
\begin{equation} \label{KS}
    \tilde{\Lambda}_D = - \dfrac{1}{24} \, \dfrac{1}{\kappa_D^2 l_s^2} \, \lim_{\sigma \to \infty} \, g(\sigma^{-1}).
\end{equation}
This equivalence matches the integral (\ref{closedstringLambdaF}) with a limit definition. It is proven in physical terms in ref.~\cite{Kutasov:1990sv}, and it assumes the absence of physical tachyons. In the mathematical literature, this identity can be shown via a generalisation of the Rankin-Selberg-Zagier technique that lies in unfolding the $\mathbb{F}$-domain integration into an $\mathbb{S}$-domain integration by taking advantage of the modular invariance of the partition function \cite{rankin1939i, rankin1939ii, selberg1940, zagier1981, Angelantonj:2010ic}, as recently reviewed by ref.~\cite{Abel:2021tyt}. 
Similarly to the case of open strings, one might worry that the cosmological constant in eq.~(\ref{KS}) could diverge when approaching the UV-region as $\sigma \to \infty$. However, we know that such a divergence must be absent due to modular invariance. We can then interpret the finiteness of eq.~\eqref{KS} as a consequence of some sort of fermion-boson cancellation, in accordance with misaligned supersymmetry.

In particular, expanding $g(\tau_2)$ in terms of $g_n$, it is possible to infer the small-$\tau_2$ behaviour
\begin{equation} \label{closedstringsupertracetool}
    \sum_n (-1)^{F_n} g_{n} \, \e^{- 4 \pi \tau_2 n} \; \overset{\tau_2 \sim 0^+}{\simeq} \; - 24 \, \kappa_D^2 l_s^2 \, \tilde{\Lambda}_D \, \tau_2^{D/2-1}.
\end{equation}
This expression motivates the definition of the regularized supertraces. Some of these are finite as a consequence of the identity in eq.~\eqref{closedstringsupertracetool} and of the finiteness of the regularized cosmological constant \cite{Dienes:1995pm}. For open strings one can formally define supertraces \cite{Cribiori:2020sct}, but they are not manifestly related to the cosmological constant in an obvious way. An interpretation is proposed in section \ref{sec: supertraces}.

\section{Open-string misaligned supersymmetry and finiteness} \label{sec: openstring-misSUSY}
For simplicity, we start by considering open strings, which is the simplest case. According to eq.~(\ref{openstringLambda}), the key fact to make sure there are no UV-divergences is that the partition function
\begin{equation}
    M_{\mathrm{D}p}(t) = \dfrac{1}{(2t)^{\frac{p+1}{2}}} M(\I t)
\end{equation}
approaches the region $t \sim 0^+$ as a positive power, which guarantees a finite cosmological constant in the absence of tachyons. This is the main topic of this section. The focus will be on the term $M(\I t) = M(\tau = \I \tau_2)$, with $t \equiv \tau_2$, which is typically expressed as a pure Dedekind $\eta$-quotient, with the power-law prefactor being the only difference between branes of different spacetime dimensions.  We will show that the misaligned symmetry in the associated state degeneracies leads to cancellation of exponential divergences in the one-loop partition function.  A remnant modular symmetry moreover ensures that all polynomial divergences cancel, leading to a finite final result.

\subsection{Setup}
For definiteness, we focus on the class of tachyon-free open-string theories where the partition function $M = M(\tau)$ is not amenable to the special Hardy-Ramanujan-Rademacher expansion discussed by ref.~\cite{sussman2017rademacher} and section \ref{subsec: sussman}, but the negative of the shifted-argument function $\tilde{M}(\tau) = M(\tau+1/2)$ is.\footnote{Generically there can also be an overall numerical positive prefactor that leads to trivial modifications of the equations below. It is immediate to include this rescaling in our results.}  An instance of this scenario is that of an anti-D$p$-brane on top of an O$p$-plane as discussed in detail in ref.~\cite{Cribiori:2020sct}. Extensions to other more complicated scenarios are immediate.

If the partition function $M(\tau)$ has the Laurent expansion
\begin{equation}
    M(\tau) = \sum_{n \in \mathbb{N}_0} (-1)^{F_n} g_n \, q^n,
\end{equation}
then the negative of the shifted-argument function $\tilde{M}(\tau)$ reads
\begin{equation}
    - \tilde{M}(\tau) = \sum_{n \in \mathbb{N}_0} (-1)^{n+1} (-1)^{F_n} g_n \, q^n \equiv  \sum_{n \in \mathbb{N}_0} a_n q^n.
\end{equation}
Employing the Hardy-Ramanujan-Rademacher expansion of ref.~\cite{sussman2017rademacher}, in the notation reviewed in section \ref{sec: misSUSY review}, the coefficients $a_n$ are found to be
\begin{equation} \label{tildeMcoefficients}
   a_n= (-1)^{n+1} (-1)^{F_n} g_n = \sum_{\alpha \in \Gamma} \dfrac{2\pi \, c_2(\alpha) \, [c_3(\alpha)]^{\frac{c_1+1}{2}}}{[24 n]^{\frac{c_1+1}{2}}} \, \dfrac{P_\alpha(n)}{\alpha} \, I_{c_1+1} \biggl[ \biggl( \dfrac{2 \pi^2}{3 \alpha^2} \, c_3(\alpha) \, n \biggr)^{\frac{1}{2}} \biggr],
\end{equation}
where we have defined the set $\Gamma = \lbrace \alpha \in \mathbb{N}: \; c_3(\alpha) > 0 \rbrace$ for brevity. Note that the terms in eq.~(\ref{tildeMcoefficients}) are only valid for $n>0$, since eq. \eqref{d(n)} does not cover the case corresponding to $n=0$. Taking all this into account, we restrict now our attention to the case $\tau=\I \tau_2$ and the function $M(\I \tau_2)$ can be expressed in the form 
\begin{equation} \label{g-definition}
        g(\tau_2) \equiv M(\I \tau_2) = (-1)^{F_0} g_0 + \sum_{\alpha \in \Gamma} \sum_{\beta=1}^{\alpha} P_\alpha(\beta) g_\alpha(\tau_2; \beta).
\end{equation}
In this expression, the terms $P_\alpha(\beta)$, with $\beta=1,\ldots , \alpha$, are the $\alpha$ different values that the periodic function $P_\alpha(n)$ can assume. Moreover, we denote by $\mathbb{N}_\alpha(\beta) = \lbrace n \in \mathbb{N} : \, n = \beta \, \mathrm{mod} \, \alpha \rbrace$ the sets of integers which satisfy $P_\alpha(n)=P_\alpha(\beta)$ for all $n \in \mathbb{N}_\alpha(\beta)$. We have also defined the functions 
\begin{equation}
    g_\alpha(\tau_2; \beta) = \sum_{n \in \mathbb{N}_\alpha(\beta)}\!\! (-1)^{n+1} \dfrac{2\pi c_2(\alpha) [c_3(\alpha)]^{\frac{c_1+1}{2}}}{\alpha [24 \, n]^{\frac{c_1+1}{2}}} I_{c_1+1} \biggl[ \biggl( \dfrac{2 \pi^2}{3 \alpha^2} c_3(\alpha) \, n \biggr)^{\frac{1}{2}} \biggr] \e^{-2 \pi \tau_2 n}.
\end{equation}
Up to this point, we have reorganized the sum over $n\in \mathbb{N}$ into $\alpha$ sums over $n \in \mathbb{N}_\alpha(\beta)$, for $\beta=1,\dots,\alpha$. For each of these sums, the quantity $P_\alpha(\beta)$ factorizes out, due to its periodicity.

It is actually convenient to make a further distinction, namely to distinguish the contributions for which $(-1)^{n+1}$ is positive from those for which it is negative. In ref.~\cite{Cribiori:2020sct}, the open-string cases of misaligned supersymmetry that were studied had only odd values of $\alpha$ contribute. 
This is also assumed here. Then, one can introduce the two sets $\mathbb{N}^\pm_\alpha(\beta) = \lbrace n \in \mathbb{N}_\alpha(\beta): \, (-1)^{n+1} = \pm 1 \rbrace$ and express the full function $g(\tau_2)$ as
\begin{equation} \label{g}
    g(\tau_2) = (-1)^{F_0} g_0 + \sum_{\alpha \in \Gamma} \sum_{\beta=1}^{\alpha} P_\alpha(\beta) \bigl( g^+_\alpha(\tau_2; \beta) - g^-_\alpha(\tau_2; \beta)),
\end{equation}
where the two definite-sign functions $g^\pm_\alpha(\tau_2; \beta)$ have been defined as
\begin{equation} \label{g^pm}
    g^\pm_\alpha(\tau_2; \beta) = \sum_{n \in \mathbb{N}^\pm_\alpha(\beta)}\!\! \dfrac{2\pi c_2(\alpha) [c_3(\alpha)]^{\frac{c_1+1}{2}}}{\alpha [24 \, n]^{\frac{c_1+1}{2}}} I_{c_1+1} \biggl[ \biggl( \dfrac{2 \pi^2}{3 \alpha^2} c_3(\alpha) \, n \biggr)^{\frac{1}{2}} \biggr] \e^{-2 \pi \tau_2 n}.
\end{equation}
Notice that for the functions $g^\pm_\alpha(\tau_2; \beta)$ the superscript sign does not relate to their effective contribution to $g(\tau_2)$ being positive or negative: this also depends on the sign of the overall term $P_\alpha(\beta)$ they are multiplied with. In the rest of this section, eqs.~(\ref{g}) and (\ref{g^pm}) will constitute the fundamental tool to discuss misaligned supersymmetry.

\subsection{Cancellation of exponential divergences} \label{subsec: open-string exponential divergences cancellations}
In order to discuss the behaviour of the function $g(\tau_2)$ in eq.~(\ref{g}), one can take advantage of the Taylor expansion of the modified Bessel function of the first kind, i.e.~\cite{AS}
\begin{equation} \label{taylorI}
    I_\delta(z) = \Bigl( \dfrac{z}{2} \Bigr)^\delta \sum_{k=0}^\infty \dfrac{\Bigl( \dfrac{z^2}{4} \Bigr)^{k}}{k! \, (\delta + k)!},
\end{equation}
where it is understood that $\delta$ is a positive integer. Thanks to this, setting $\delta = c_1 + 1$, the functions in eq.~(\ref{g^pm}) can be expressed as
\begin{equation} \label{g^pm2}
    g^\pm_{\alpha}(\tau_2; \beta) = 2\pi c_2(\alpha)\, \alpha^{c_1} \sum_{n \in \mathbb{N}^\pm_\alpha(\beta)} \sum_{k=0}^\infty \; \biggl[\dfrac{\pi}{12} \dfrac{c_3(\alpha)}{\alpha^2} \biggr]^{c_1 + k +1} \dfrac{(2 \pi n)^{k}}{k! \, (c_1 + k +1)!} \, \e^{-2 \pi \tau_2 n}.
\end{equation}
This expression makes it possible to study the region $\tau_2 \sim 0^+$ in quite a fruitful way. In what follows, we will consider a finite $\tau_2 > 0$ in order to carry out the calculations with the infinite sums. Then, we will assess the behaviour of the functions of interest in the limit $\tau_2\to 0^+$.

Because the elements in the infinite summations over $k$ and $n$ are positive-definite, the order of the two summations in eq.~(\ref{g^pm2}) can be interchanged. The sum for $n \in \mathbb{N}^\pm_\alpha(\beta)$ can be rearranged by observing that its elements can be written as $\smash{n = m^\pm_\alpha(\beta) \, \mathrm{mod} \, \gamma_\alpha}$, where $\smash{m^\pm_\alpha}(\beta)$ is an integer depending on $\alpha$ and $\beta$ and $\gamma_\alpha = \mathrm{lcm} \, (2, \alpha) = 2 \alpha$, with $\alpha$ assumed to be odd. Note that $\smash{m^\pm_\alpha(\beta)}$ is by definition the smallest element in the set $\smash{\mathbb{N}^\pm_\alpha(\beta)}$, and it is generally not corresponding to $\beta$. For example, $\smash{m^+_\alpha(\beta)}$ is the smallest positive odd (since $(-1)^{n+1}\equiv 1$) integer equal to $\beta  \, \mathrm{mod} \,\alpha$. Since we assume $\alpha$ to be odd, if $\beta$ is odd too we have $\smash{m^+_\alpha(\beta)} = \beta$, while if $\beta$ is even  $\smash{m^+_\alpha(\beta)} = \beta+\alpha$, which is odd. A similar reasoning applies to $\smash{m^-_\alpha(\beta)}$. In general we can write
\begin{equation}\label{m_beta}
    m^\pm_\alpha(\beta) = \beta + \dfrac{(1 \pm (-1)^\beta)}{2} \alpha.
\end{equation}
This will be helpful later on, but for now it can be left unexpanded. With this parametrisation, the summation over $n$ can be performed in terms of the geometric series, resulting in\footnote{To evaluate the series by writing $\smash{(2 \pi n)^k \e^{- 2 \pi \tau_2 n} = (-1)^k (\de / \de \tau_2)^k \, \e^{-2 \pi \tau_2 n}}$, one must invert the order of the differentiation with respect to $\smash{k}$ and of the summation over $\smash{n}$. For a series of functions $\smash{f_n(x)}$, if their series $\smash{f(x) = \sum_{n \in \mathbb{N}} f_n(x)}$ is convergent and if the series of their derivatives $\smash{\sum_{n \in \mathbb{N}} f'_n(x)}$ is uniformly convergent, then the identity holds $\smash{f'(x) = \sum_{n \in \mathbb{N}} f'_n(x)}$ (see eq.~(0.307) in ref.~\cite{GR}). For the case at hand, the series is not convergent in the region $\smash{\tau_2 \sim 0^+}$, as shown by the term  $\smash{1/\tau_2}$, so one should remove this and consider the leftover sum. This is indeed our working assumption. An alternative way to compute the required series rigorously is to make use of the results for the arithmetico-geometric sum (see eq.~(0.113) in ref.~\cite{GR}).}
\begin{equation}
    \begin{split}
        \sum_{n \in \mathbb{N}^\pm_\alpha(\beta)} (2 \pi n)^{k} \, \e^{-2 \pi \tau_2 n} & = \sum_{l=0}^\infty [2\pi (m^\pm_\alpha(\beta) + l \gamma_\alpha)]^k \, \e^{-2 \pi \tau_2 \left[ m^\pm_\alpha(\beta) + l \gamma_\alpha \right]} \\
        & = (-1)^k \dfrac{\de^k}{\de \tau_2^k} \sum_{l=0}^\infty \e^{-2 \pi \tau_2 \left[ m^\pm_\alpha(\beta) + l \gamma_\alpha \right]} \\
        & = (-1)^k \dfrac{\de^k}{\de \tau_2^k} \biggl[ \dfrac{\e^{2 \pi \left[ \gamma_\alpha - m^\pm_\alpha(\beta) \right] \tau_2}}{\e^{2 \pi \gamma_\alpha \tau_2} - 1} \biggr].
    \end{split}
\end{equation}
In this way, to finally explore the region where $\tau_2 \sim 0^+$, it is sufficient to Taylor-expand the leftover order-$k$ derivative. From the expansion (we refer the reader to the appendix \ref{app:specialfunctions} for the notation)
\begin{equation}
   \dfrac{\e^{2 \pi \left[ \gamma_\alpha - m^\pm_\alpha(\beta) \right] \tau_2}}{\e^{2 \pi \gamma_\alpha \tau_2} - 1}  = \dfrac{1}{2 \pi \gamma_\alpha} \dfrac{1}{\tau_2} + \dfrac{ \gamma_\alpha-2m^\pm_\alpha(\beta)}{2 \gamma_\alpha} + O(\tau_2;0),
\end{equation}
we learn that the function to be differentiated $k$ times at leading order is $1 / (2 \pi \gamma_\alpha \tau_2)$. It should be noted that this is the only $\beta$- and $(\pm)$-independent term; the leftover power series depends on $\beta$ and the $(\pm)$-sign via the terms $m^\pm_\alpha(\beta)$. In more detail, one obtains 
\begin{equation} \label{taylor}
    (-1)^k \dfrac{\de^k}{\de \tau_2^k} \biggl[ \dfrac{\e^{2 \pi \left[ \gamma_\alpha - m^\pm_\alpha(\beta) \right] \tau_2}}{\e^{2 \pi \gamma_\alpha \tau_2} - 1} \biggr] = \dfrac{1}{2 \pi \gamma_\alpha} \dfrac{k!}{\tau_2^{1+k}} + \sum_{l=0}^\infty f_l(k, m^\pm_\alpha(\beta)) \tau_2^l,
\end{equation}
where $\smash{f_l(k, m^\pm_\alpha(\beta))}$ are constants not depending on $\tau_2$ that will be discussed later on (see eq.~(\ref{f_l}) for their explicit expression). Therefore, we have been able to perform the sum over $n$ in eq.~(\ref{g^pm}). Thanks to the expansion of eq.~\eqref{taylor}, the original function $g^\pm_{\alpha}(\tau_2; \beta)$ appearing in eq.~(\ref{g^pm}), and rearranged into a different form in eq.~(\ref{g^pm2}), can now be written as 
\begin{equation} \label{g^pm3}
    g^\pm_{\alpha}(\tau_2; \beta) = \dfrac{\alpha^{c_1}}{\tau_2} \dfrac{c_2(\alpha)}{\gamma_\alpha} \biggl[\dfrac{\pi}{12} \dfrac{c_3(\alpha)}{\alpha^2} \biggr]^{c_1+1} \sum_{k=0}^\infty \; \dfrac{\biggl[ \dfrac{\pi}{12} \dfrac{c_3(\alpha)}{\alpha^2} \dfrac{1}{\tau_2} \biggr]^{k}}{(c_1 + k+1)!} + \Delta g^\pm_{\alpha}(\tau_2; \beta),
\end{equation}
where, according to eq.~(\ref{taylor}), the remainder is
\begin{equation} \label{Deltag_alpha^pm}
    \Delta g^\pm_{\alpha}(\tau_2; \beta) = 2\pi c_2(\alpha)\,\alpha^{c_1} \sum_{k=0}^\infty \sum_{l=0}^\infty \dfrac{\biggl[\dfrac{\pi}{12} \dfrac{c_3(\alpha)}{\alpha^2} \biggr]^{c_1 + k +1\!\!\!\!}}{k! \, (c_1 + k +1)!} f_l(k, m^\pm_\alpha(\beta)) \tau_2^l.
\end{equation}
So, eq.~(\ref{g^pm3}) contains a singular part as $\tau_2 \sim 0^+$ and a power-series remainder. As anticipated above, the key difference among these two terms consists in the fact that only the power series has a dependence on $\beta$ and the $(\pm)$-sign. In the singular part, one can recognize the leftover sum to be
\begin{equation}
\label{expdiv}
    \sum_{k=0}^\infty \; \dfrac{\biggl[ \dfrac{\pi}{12} \dfrac{c_3(\alpha)}{\alpha^2} \dfrac{1}{\tau_2} \biggr]^{k}}{(c_1 + k+1)!} = \dfrac{\e^{\frac{\pi}{12} \frac{c_3(\alpha)}{\alpha^2} \frac{1}{\tau_2}}}{\biggl[\dfrac{\pi}{12} \dfrac{c_3(\alpha)}{\alpha^2} \dfrac{1}{\tau_2}\biggr]^{c_1+1}} \Biggl[ 1 - \dfrac{1}{c_1!} \Gamma \biggl[c_1+1,\dfrac{\pi}{12} \dfrac{c_3(\alpha)}{\alpha^2} \dfrac{1}{\tau_2}\biggr] \Biggr],
\end{equation}
where $\Gamma(\nu,z)$ is the incomplete $\Gamma$-function.
In the region $\tau_2 \sim 0^+$, the incomplete $\Gamma$-function can also be expanded to write
\begin{equation} \label{incomplete Gamma-function expansion}
    \sum_{k=0}^\infty \dfrac{\biggl[ \dfrac{\pi}{12} \dfrac{c_3(\alpha)}{\alpha^2} \dfrac{1}{\tau_2} \biggr]^{k}}{(c_1+1 + k)!} = \dfrac{\e^{\frac{\pi}{12} \frac{c_3(\alpha)}{\alpha^2} \frac{1}{\tau_2}}}{\biggl[\dfrac{\pi}{12} \dfrac{c_3(\alpha)}{\alpha^2} \dfrac{1}{\tau_2}\biggr]^{c_1+1}} - \dfrac{1}{c_1!} \dfrac{\tau_2}{\biggl[\dfrac{\pi}{12} \dfrac{c_3(\alpha)}{\alpha^2}\biggr]} + O(\tau_2;0)^{2}.
\end{equation}
One can eventually conclude that the function $g^\pm_{\alpha}(\tau_2; \beta)$ around the point $\tau_2 \sim 0^+$ reads
\begin{equation} \label{g^pm asymptotic}
    g^\pm_{\alpha}(\tau_2; \beta) \overset{\tau_2 \sim 0^+}{\simeq} \dfrac{c_2(\alpha)}{\gamma_\alpha} \,\alpha^{c_1}\,\tau_2^{c_1} \, \e^{\frac{\pi}{12} \frac{c_3(\alpha)}{\alpha^2} \frac{1}{\tau_2}} + r(\alpha,\tau_2) + \Delta g^\pm_{\alpha}(\tau_2; \beta),
\end{equation}
where the exponential term comes from the leading divergent term in eq.~(\ref{incomplete Gamma-function expansion}), with an associated finite remainder
\begin{equation} \label{remainder}
    r(\alpha,\tau_2) = - \dfrac{1}{c_1!} \dfrac{c_2(\alpha)}{\gamma_\alpha}\,\alpha^{c_1}\, \biggl(\dfrac{\pi}{12} \dfrac{c_3(\alpha)}{\alpha^2}\biggr)^{c_1} + O(\tau_2;0),
\end{equation}
and $\Delta g^\pm_{\alpha}(\tau_2; \beta)$ is the polynomial term defined in eq.~(\ref{Deltag_alpha^pm}). The functions in eq.~(\ref{g^pm asymptotic}) obviously diverge for $\tau_2 \to 0$ sector by sector due to the exponential of $1/\tau_2$. However, the complete physical information relating to the one-loop cosmological constant is contained in the sum over sectors in the function $g(\tau_2)$ defined in eq.~(\ref{g}), and in this sum the singular part is automatically cancelled out by the fermion-boson oscillation appearing therein at order $\alpha=1$ and by the Hardy-Ramanujan-Rademacher-expansion property $\sum_{\beta=1}^{\alpha} P_\alpha (\beta) = 0$ for higher orders $\alpha>1$. This is true for all the contributions coming from the term scaling as $1/\tau_2^{k+1}$ in the expansion of eq.~(\ref{taylor}), i.e. not only for the leading term in eq.~\eqref{g^pm asymptotic} but also for the remainder in eq.~(\ref{remainder}), since they all are independent of $\beta$ and the $(\pm)$-sign. Notice that, as an alternative, one may also still explain the subleading-order cancellations in view of boson-fermion cancellations, in a similar way as for the leading-order terms. We find it interesting to emphasise that for subleading orders this is not necessary, and moreover the visualisation of the cancellations in view of the property $\sum_{\beta=1}^{\alpha} P_\alpha (\beta) = 0$ is instrumental in elucidating the closed-string analysis (see section \ref{sec: closedstring-misSUSY}). All cancellations find an intuitive interpretation in the anti-D$p$-brane/O$p$-plane example represented in fig.~\ref{corrections to anti-D/O plot}. Finally, since all the $\beta$-independent terms appearing in $g_\alpha^\pm(\tau_2; \beta)$ cancel, eq.~(\ref{g}) can be simply written as
\begin{equation} \label{g2}
    g(\tau_2) = (-1)^{F_0} g_0 + \sum_{\alpha \in \Gamma} \sum_{\beta=1}^{\alpha} P_\alpha(\beta) \bigl[ \Delta g^+_\alpha(\tau_2; \beta) - \Delta g^-_\alpha(\tau_2; \beta) \bigr].
\end{equation}
Remarkably, this is just a constant term plus a power-series difference. Therefore, we proved that all of the exponentially divergent contributions to the one-loop cosmological constant coming from the first part of eq.~\eqref{g^pm3} (namely those contained in eq.~\eqref{expdiv}) cancel out when summing over all of the sectors of the theory, leaving at most a polynomial dependence on $\tau_2$. The only thing that matters to reach this result is that all of these singular contributions to $g^\pm_{\alpha}(\tau_2; \beta)$ are identical for a given $\alpha$ (i.e.~they are independent of $\beta$ and of the $(\pm)$-sign) and therefore cancel out when summing over the sectors labelled by $\beta$ and/or when taking into account the difference between positive and negative terms.

To summarise, we have shown how the exponential divergences appearing in the open-string one-loop cosmological constant eq.~(\ref{openstringLambda}) cancel, thanks to the misaligned supersymmetry in the spectrum of state degeneracies.  This result follows the cancellations found in the sector-averages defined in refs.~\cite{Dienes:1994np, Cribiori:2020sct}.  To compute the sector-averages, it was necessary to define sector degeneracies $a_n(\alpha)$ for discrete towers $n$ at each order $\alpha$ in the Hardy-Ramanujan-Rademacher-expansion.  Further, at each order $\alpha$, different subsectors labelled by $\beta=1,\dots,\alpha$ were introduced, whose degeneracies could be extrapolated to the envelope functions $\Phi_\beta (n; \alpha)$, defined for continuous $n \in \mathbb{R}^+$.  The envelope functions could then be summed into the sector-average, and the cancellations observed.  Although the cancellations seemed remarkable, the physical meaning of (sub)sectors, envelope functions and sector-average was unclear. We have shown above that the same cancellations actually occur directly in the partition function $M(\I \tau_2)=g(\tau_2)$ and thus in physical quantities like the one-loop cosmological constant.

\subsection{Cancellation of polynomial divergences}

In order to claim finiteness of the one-loop cosmological constant, the leftover polynomial terms in eq.~(\ref{g2}) need to be studied carefully as $\tau_2 \sim 0^+$. Indeed, although we have proven that exponential divergences are absent, the integral defining the cosmological constant may still be singular as a power-law. In general one can write
\begin{equation} \label{g3}
    g(\tau_2) = (-1)^{F_0} g_0 + \Delta g^+ (\tau_2) - \Delta g^- (\tau_2) = (-1)^{F_0} g_0 + \sum_{l=0}^{\infty} b_l \tau_2^l,
\end{equation}
where the $\tau_2$-dependence comes from the difference of two simple power series
\begin{equation} \label{Deltag^pm}
    \Delta g^\pm (\tau_2) = \sum_{\alpha \in \Gamma} \sum_{\beta=1}^{\alpha} P_\alpha(\beta) \Delta g^\pm_{\alpha}(\tau_2; \beta) = \sum_{l=0}^\infty b^\pm_l \tau_2^l,
\end{equation}
with the definition $b_l = b^+_l - b^-_l$. On the other hand, the constant term is $g(0) = (-1)^{F_0} g_0 + b_0$. A few manipulations, summarised in appendix \ref{app: bl-derivation}, allow one to determine an explicit expression for the coefficients of the power series $\Delta g^\pm (\tau_2)$. In fact, one can show that the power-series coefficients read
\begin{equation} \label{b}
    b_l = \dfrac{\pi}{l!} \sum_{\alpha \in \Gamma} \dfrac{\alpha^{l-1} c_2(\alpha)}{(2\pi)^{c_1-l+1}} \sum_{k=0}^\infty \dfrac{\biggl[\dfrac{\pi^2}{6} \dfrac{c_3(\alpha)}{\alpha} \biggr]^{c_1+k+1\!\!\!\!}}{k! \, (c_1+k + 1)!} \, \sum_{r=0}^{\alpha-1} (-1)^{k+r} P_\alpha(-r) E_{k+l} \Bigl( \dfrac{r}{\alpha} \Bigr).
\end{equation}
This represents the coefficient of the order-$l$ term in the power series $\Delta g(\tau_2)$ for a generic open-string model where only odd values of $\alpha$ appear in the Hardy-Ramanujan-Rademacher expansion.  Although it is difficult to further reduce the expression (\ref{b}) directly,\footnote{The complication in eq.~(\ref{b}) lies in the form of the Kloosterman-like term $P_\alpha(-r)$, which is hard to deal with analytically.} we will now do so indirectly by using some simple observations on the Dedekind $\eta$-function due to Zagier \cite{ZeidlerZagier}.

Let the partition function be a Dedekind $\eta$-quotient
\begin{equation}
    M(\tau) = \xi \, \prod_{m=1}^\infty \bigl[ \eta(m \tau) \bigr]^{\delta_m},
\end{equation}
for some constant $\xi$. By exploiting the modular properties of the string partition function, it is possible to determine the behaviour of this function on the imaginary axis $\tau_1=0$ as $\tau_2 \sim 0^+$. In fact, under the generating $\mathrm{S}$-transformation $S(\tau) = - 1 / \tau$ of the modular group $\mathrm{PSL}_2(\mathbb{Z})$, the Dedekind $\eta$-function transforms as $\eta(-1/\tau) = \sqrt{- \I \tau} \, \eta (\tau)$, so, restricting to the imaginary axis $\tau = \I\tau_2$, one can write 
\begin{equation}
    \eta \Bigl( \dfrac{\I}{\tau_2} \Bigr) = \sqrt{\tau_2} \, \eta (\I \tau_2).
\end{equation}
The Dedekind $\eta$-function can be written as $\eta(\I t) = \e^{- \frac{\pi t}{12}} \prod_{m=1}^\infty (1 - \e^{- 2 \pi m t} )$, which gives
\begin{equation}\label{asymptotic_eta2}
    \mathrm{ln} \, \eta (\I t) = - \dfrac{\pi t}{12} + \sum_{m=1}^\infty \mathrm{ln} \, (1 - \e^{- 2 \pi m t} ) = - \dfrac{\pi t}{12} + O(\e^{- 2 \pi t};\infty).
\end{equation}
So, combining the $\mathrm{S}$-transformation relation and the limit as $1/\tau_2 \sim \infty$, one concludes that, in the region where $\tau_2 \sim 0^+$, the Dedekind $\eta$-function behaves as (see appendix \ref{subapp:asymptoticeta} for more details) 
\begin{equation} \label{asymptotic_eta}
    \eta(\I \tau_2) \overset{\tau_2 \sim 0^+}{\simeq} \tau_2^{-\frac{1}{2}} \, \e^{-\frac{\pi}{12 \tau_2}}.
\end{equation}
Therefore, by defining the coefficients 
\begin{equation}
    s = \prod_{m=1}^\infty m^{\delta_m}, \qquad c_1 = - \dfrac{1}{2} \sum_{m=1}^\infty \delta_m, \qquad c_4 = - \sum_{m=1}^\infty \frac{\delta_m}{m},
\end{equation}
one can simply write the asymptotic behaviour of the open-string partition function as
\begin{equation} \label{asymptotic_M}
    M(\I \tau_2) \overset{\tau_2 \sim 0^+}{\simeq} \xi \, s^{-\frac{1}{2}} \, \tau_2^{c_1} \, \e^{\frac{\pi c_4}{12 \tau_2}}.
\end{equation}
In the absence of an exponential divergence, i.e.~for $c_4=0$, which we assume to be true in subsection \ref{subsec: open-string exponential divergences cancellations} and verify for all the explicit examples we consider, this provides a direct way to compute the power-series coefficients (\ref{b}) appearing in the expansion of eq.~(\ref{g3}). Assuming $c_1$ to be an integer, which is also verified in our examples, eq.~\eqref{asymptotic_M} indicates that the constant term and the first $c_1-1$ coefficients are zero and that the first non-zero one is $b_{c_1}$, i.e.
\begin{subequations} \label{b-coefficient values}
\begin{align}
    & (-1)^{F_0} g_0 + b_{0} = b_1 = \dots = b_{c_1-1} = 0, \\
    & b_{c_1} = \xi \, s^{-\frac{1}{2}}.
\end{align}
\end{subequations}
Note that not only can we easily find this leading polynomial term, but we can actually also show that all the coefficients $b_l$ except $b_{c_1}$ are zero. Indeed, when using eq.~\eqref{asymptotic_eta2} in eq.~\eqref{asymptotic_eta}, we find
\begin{equation} 
    \eta(\I \tau_2) = \tau_2^{-\frac{1}{2}} \, \e^{-\frac{\pi}{12 \tau_2}} \Bigl[1 + O\Bigl(\e^{-\frac{2\pi}{\tau_2}},0 \Bigr)\Bigr].
\end{equation}
This then means that eq.~\eqref{asymptotic_M} is only corrected by terms that are exponentially suppressed compared to the leading polynomial term. Therefore, we find that $b_{c_1} \tau_2^{c_1}$ is the only non-zero polynomial term. From the discussion in section \ref{subsec: openstringLambda}, we see then that the cosmological constant of a D$p$-brane theory is not divergent if $c_1>(p+1)/2$.

It is useful to illustrate these general results with an explicit example. For an anti-D$p$-brane on top of an O$p$-plane, the $p$-independent part of the partition function is
\begin{equation}
    - \dfrac{1}{8} \, M(\tau) = \dfrac{\eta^{16}(\tau) \, \eta^{16}(4 \tau)}{\eta^{40}(2\tau)}.\label{eq:Mopen}
\end{equation}
For this, one finds $s = 1/256$, $c_1=4$, and the exponential disappears as $c_4=0$, which means
\begin{equation}
    g(\tau_2) = M(\I \tau_2) = - 128 \tau_2^4 + O\Bigl(\e^{-\frac{2\pi}{\tau_2}},0 \Bigr).\label{eq:Mopenexpansion}
\end{equation}
So, we find the expected cancellation of divergent terms and we can explicitly determine the full power-law behaviour, finding only one non-zero term. Note that for the open string there is no analogue of the Kutasov-Seiberg formula and, in order to determine the cosmological constant, we have to do the integral in eq.~\eqref{openstringLambda}. The above cancellations and power-law behaviour ensure the finiteness of the integral for small $\tau_2$, whilst the absence of physical tachyons ensures finitess for large $\tau_2$. One can compute the finite value of the integral numerically \cite{Cribiori:2020sct}.

To summarise, whilst we showed explicitly how misaligned state degeneracies lead to a cancellation of exponential divergences in the open-string one-loop cosmological constant, we used modular invariance to prove that the polynomial divergences cancel.  Although it has not been possible to show it directly, modular invariance must constrain the state degeneracies in such a way as to ensure these cancellations, leading to the mathematical identities $b_l=0$ for $l\neq c_1$, with $b_l$ defined in eq. (\ref{b}).
It is also interesting to note that, whilst the behaviour in eq.~(\ref{asymptotic_M}) has been explained as a consequence of the $\mathrm{PSL}_2(\mathbb{Z})$-properties of the Dedekind $\eta$-function, it can also be inferred from simpler considerations in mathematical analysis that are in fact independent of modular invariance \cite{ZeidlerZagier}. Details about both methods are in appendix \ref{subapp:asymptoticeta}.

\section{Closed-string misaligned supersymmetry and finiteness} \label{sec: closedstring-misSUSY}

To describe misaligned supersymmetry for closed strings, the fundamental object to discuss is the function $g(\tau_2)$ defined  in eq.~(\ref{g(tau2)}). The Kutasov-Seiberg identity (\ref{KS}) directly relates the function $g(\tau_2)$ to the one-loop cosmological constant and the latter is finite so long as $g(\tau_2)$ approaches the region $\tau_2 \sim 0^+$ as a constant. The discussion is more complicated compared to the case of open strings since the partition function is the product of a right- and a left-moving sector, but the analysis follows the same pattern. For this reason, we will mainly outline the relevant steps and differences with respect to the analysis in the previous section.

\subsection{Setup} \label{subsec: closed-string setup}
Let the closed-string partition function be of the form $\smash{Z(\tau, \overline{\tau}) = \tau_2^{1-D/2} R(\tau) \overline{L}(\overline{\tau})}$, where the terms $\smash{R(\tau) = q^{-n_0^R} \sum_{n=0}^\infty a^R_n q^n}$ and $\smash{L(\tau) = q^{-n_0^L} \sum_{n=0}^\infty a^L_n q^n}$ are the right- and left-moving contributions, respectively, with $\smash{q = \e^{2 \pi \I\tau}}$. More generally, the closed-string partition function can be the sum of several such terms, i.e. $Z(\tau,\overline{\tau}) = \tau_2^{1-D/2} \sum_\sigma Z_\sigma(\tau, \overline{\tau})$, with $Z_\sigma(\tau, \overline{\tau}) = R_\sigma(\tau) \overline{L}_\sigma(\overline{\tau})$, in which case our discussion of exponential divergences below may be applied to each term $Z_\sigma(\tau,\overline{\tau})$ individually. This is the case for example for the heterotic $\mathrm{SO}(16) \!\times\! \mathrm{SO}(16)$-theory in ref.~\cite{Cribiori:2020sct}. It should be pointed out that it is conceivable that there also may be models in which the cancellations happen between different terms, and this would require an adaptation of the procedure discussed below. Also notice that for simplicity here we consider the case where $n \in \mathbb{N}_0$; terms with $n \in \mathbb{N}_0/2$ can be studied similarly after a rescaling of the variable $\tau'=2\tau$. The constant terms $n_0^R$ and $n_0^L$ are assumed to be integer, which can also follow from a rescaling. Then, one can write
\begin{equation}
    g(\tau_2) = \tau_2^{1-D/2} \sum_{n = - n_0}^\infty (-1)^{F_n} g_n \, \e^{-4 \pi \tau_2 n},
\end{equation}
where, defining $n_0 = \mathrm{min} \, (n_0^R,  n_0^L)$, the net physical degeneracies are
\begin{equation}
    (-1)^{F_n} g_n = a^R_{n + n_0^R} \, \overline{a}^L_{n + n_0^L}.
\end{equation}
If both the functions $R(\tau)$ and $L(\tau)$ are Dedekind $\eta$-quotients that are amenable to the special Hardy-Ramanujan-Rademacher-expansion analysed in ref.~\cite{sussman2017rademacher}, then it is possible to express the Laurent coefficients $a^R_{n + n_0^R}$ and $\overline{a}^L_{n + n_0^L}$ as simplified Hardy-Ramanujan-Rademacher sums, for $n>0$. In fact, it is possible to write
\begin{equation}
    g(\tau_2) = \tau_2^{1-D/2} \bigl[ h_0(\tau_2) + h(\tau_2) \bigr],
\end{equation}
where $h_0(\tau_2)$ represents the sum restricted to coefficients not given by the Hardy-Ramanujan-Rademacher-expansion and $h(\tau_2)$ stands for the remaining infinite series, i.e.
\begin{subequations}
\begin{align}
    h_0(\tau_2) & = \sum_{n=0}^{n_0} (-1)^{F_{-n}} g_{-n} \, \e^{4 \pi \tau_2 n}, \\
    h(\tau_2) & = \sum_{n \in \mathbb{N}} \sum_{\alpha \in \Gamma_R} \sum_{\beta \in \Gamma_L} P^R_\alpha (n+n_0^R) \overline{P}^L_\beta (n+n_0^L) f^R_{n+n_0^R}(\alpha) \overline{f}^L_{n+n_0^L} (\beta) \, \e^{-4 \pi \tau_2 n}. \label{h(tau2)}
\end{align}
\end{subequations}
Here, we have defined the two sets containing the contributions to the coefficients, i.e. $\Gamma_R = \lbrace \alpha \in \mathbb{N}: c_3^R(\alpha)>0 \rbrace$ and $\Gamma_L = \lbrace \beta \in \mathbb{N}: c_3^L(\beta)>0 \rbrace$, and the functions $f^R_n(\alpha)$ and $f^L_n(\beta)$ contain the rest of the Hardy-Ramanujan-Rademacher-expansions factors aside from the $P$-functions. We can see from eq.~\eqref{closedstringLambdaF} that we find a diverging cosmological constant from the term $h_0(\tau_2)$ if and only if $n_0 \neq 0$. In this case we have physical tachyons in the spectrum and therefore no stable vacuum around which we can study the theory. For such cases the Kutasov-Seiberg identity in eq.~\eqref{KS} is not applicable and we will therefore restrict ourselves to theories with $n_0=0$, which implies $h_0(\tau_2)= (-1)^{F_{0}} g_{0}$.

Because of the periodicity of the functions $P^R_\alpha(n)$ and $P^L_\beta(n)$, given the index $\ell = 1, \dots, \mathrm{lcm} \, (\alpha, \beta)$, with the dependence on $\alpha$ and $\beta$ on its range being left implicit for brevity, we can rearrange the infinite sum over $n$ in $h(\tau_2)$ by writing\footnote{When summing over $n$, we have to be careful in exploiting properly the periodicity of the $P$-functions. The correct strategy is explained in ref.~\cite{Cribiori:2020sct}. We split the sum over $n$ into $\ell = 1,\dots, \mathrm{lcm}(\alpha,\beta)$ contributions, in front of which the $P$-functions factorize. Then, within each of these contributions we have to sum the $f_n(\alpha)$ over all of the possible values of $n$ associated to the fixed $\ell$, namely those for which $n = \ell \, \mathrm{mod} \, \mathrm{lcm}(\alpha,\beta)$. This is needed since we want to sum over all $n$ such that $P^R_{\alpha} (n+n_0^R) \bar P^L_{\beta} (n+n_0^L) =P^R_{\alpha} (\ell+n_0^R) \bar P^L_{\beta} (\ell+n_0^L) $, for a fixed $\ell$. In fact, the product $P^R_{\alpha} (n+n_0^R) \bar P^L_{\beta} (n+n_0^L)$ is unchanged by an $\mathrm{lcm}(\alpha,\beta)$-step.}

\begin{equation} \label{gclosed}
    h(\tau_2) = \sum_{\alpha \in \Gamma_R} \sum_{\beta \in \Gamma_L} \!\!\sum_{\ell=1}^{\mathrm{lcm}(\alpha,\beta)}\!\! P^R_\alpha (\ell+n_0^R) \overline{P}^L_\beta (\ell+n_0^L) \, h_{\alpha \beta} (\tau_2; \ell),
\end{equation}
where we have defined the functions
\begin{equation} \label{g_abclosed}
    h_{\alpha \beta} (\tau_2; \ell) = \sum_{n \in \mathbb{N}_{\alpha \beta}(\ell)}\!\! f^R_{n+n_0^R}(\alpha) \overline{f}^L_{n+n_0^L} (\beta) \, \e^{-4 \pi \tau_2 n},
\end{equation}
with the sets $\mathbb{N}_{\alpha \beta} (\ell) = \lbrace n \in \mathbb{N} : \, n = \ell \, \mathrm{mod} \, \mathrm{lcm}(\alpha,\beta) \rbrace$ being defined in such a way that the condition $ P^R_\alpha (n + n_0^R) \overline{P}^L_\beta (n + n_0^L) = P^R_\alpha (\ell + n_0^R) \overline{P}^L_\beta (\ell + n_0^L)$ holds for all $n\in \mathbb{N}_{\alpha \beta} (\ell)$. In a straightforward calculation, analogous to the open-string one discussed above, one can show that if the functions $h_{\alpha \beta} (\tau_2; \ell)$ have a divergent exponential term which is independent of $\ell$, then the vanishing of the pure $P$-function combinations in one sector is enough to conclude that such exponential divergences cancel out. This is discussed below.

\subsection{Cancellation of exponential divergences} \label{ssec:closedexpcancel}
By making use of the explicit form of the functions $f^R_n(\alpha)$ and $f^L_n(\beta)$, and thanks to the Taylor expansion of the Bessel function, one can write the functions in eq.~(\ref{g_abclosed}) as

\begin{equation}
    \begin{split}
        h_{\alpha \beta} (\tau_2; \ell) = \dfrac{4\pi^2 c^R_2(\alpha) c^L_2(\beta)}{\alpha^{-c^R_1} \beta^{-c_1^L}} & \sum_{a=0}^\infty \sum_{b=0}^\infty \biggl[ \dfrac{\pi}{12} \dfrac{c^R_3(\alpha)}{\alpha^2} \biggr]^{c_1^R\!+\!a+1} \biggl[ \dfrac{\pi}{12} \dfrac{c^L_3(\beta)}{\beta^2} \biggr]^{c^L_1\!+\!b+1} \\
        & \times \!\!\!\!\!\!\sum_{n \in \mathbb{N}_{\alpha \beta}(\ell)}\!\! \dfrac{(2 \pi n)^{a+b} \,  \e^{-4 \pi \tau_2 n}}{a! b! (c_1^R\!+\!a+1)! (c_1^L\!+\!b+1)!}.
    \end{split}
\end{equation}
Defining the step $\gamma_{\alpha \beta} = \mathrm{lcm} \, (\alpha, \beta)$, according with the definition of the sets $\mathbb{N}_{\alpha \beta} (\ell)$ above we can write
\begin{equation}
    \begin{split}
        \sum_{n \in \mathbb{N}_{\alpha \beta}(\ell)}\!\!\!\! (2 \pi n)^{a+b} \,  \e^{-4 \pi \tau_2 n} & = \sum_{k = 0}^\infty [2 \pi (\ell + k \gamma_{\alpha \beta})]^{a+b} \, \e^{-4 \pi \tau_2 (\ell + k \gamma_{\alpha \beta})} \\
        & = \biggl(\! - \dfrac{1}{2} \dfrac{\de}{\de \tau_2} \!\biggr)^{\! a+b} \dfrac{\e^{4 \pi (\gamma_{\alpha \beta} - \ell) \tau_2}}{\e^{4 \pi \gamma_{\alpha \beta} \tau_2} - 1}.
    \end{split}
\end{equation}
So the Bernoulli polynomials appear again, enabling one to write in general
\begin{equation}
    \begin{split}
        & \biggl(\! - \dfrac{1}{2} \dfrac{\de}{\de \tau_2} \!\biggr)^{\! r} \dfrac{\e^{4 \pi (\gamma_{\alpha \beta} - \ell) \tau_2}}{\e^{4 \pi \gamma_{\alpha \beta} \tau_2} - 1} = \\
        = \, & \dfrac{1}{4 \pi \gamma_{\alpha \beta} \tau_2} \dfrac{r!}{(2 \tau_2)^{r}} +  \sum_{m=0}^\infty B_{m+r+1} \biggl( \dfrac{\ell}{\gamma_{\alpha \beta}} \biggr) \dfrac{(-1)^{m+1} (2 \pi \gamma_{\alpha \beta})^{m+r}(2 \tau_2)^{m}}{(m+r+1) \, m!},
    \end{split}
\end{equation}
which eventually means
\begin{equation} \label{taylor2}
    \begin{split}
        & \sum_{n \in \mathbb{N}_{\alpha \beta}(\ell)}\!\!\!\! (2 \pi n)^{a+b} \,  \e^{-4 \pi \tau_2 n} = \\
        = \, & \dfrac{1}{4 \pi \gamma_{\alpha \beta} \tau_2} \dfrac{(a+b)!}{(2 \tau_2)^{a+b}} +  \sum_{m=0}^\infty B_{m+a+b+1} \biggl( \dfrac{\ell}{\gamma_{\alpha \beta}} \biggr) \dfrac{(-1)^{m+1} (2 \pi \gamma_{\alpha \beta})^{a+b+m} (2\tau_2)^{m}}{(m+a+b+1) \, m!}.
    \end{split}
\end{equation}
This formally looks the same as for open strings, as expected. In particular, the first term could again give rise to exponential divergences, once it is resummed into an incomplete $\Gamma$-function. However, as in the open-string case, this divergent term is manifestly independent of $\ell$ and therefore, in the full expression of $h(\tau_2)$ in eq.~(\ref{gclosed}), one can immediately make use of this to perform the sum over such $\ell$. For instance, for $\beta>\alpha$, this sum gives 
\begin{equation} \label{closed-string cancellation}
    \begin{split}
        \sum_{\ell=1}^{\mathrm{lcm}(\alpha,\beta)}\!\! P^R_\alpha (\ell+n_0^R) \overline{P}^L_\beta (\ell+n_0^L) & = \sum_{k_\alpha=1}^\alpha \! \sum_{m=0}^{\frac{\beta}{\mathrm{gcd}(\alpha, \beta)}-1} \! P^R_\alpha (k_\alpha + m \alpha + n_0^R) \overline{P}^L_\beta (k_\alpha+m\alpha+n_0^L) \\
        & = \sum_{k_\alpha=1}^\alpha P^R_\alpha (k_\alpha + n_0^R) \left[ \sum_{m=0}^{\frac{\beta}{\mathrm{gcd}(\alpha, \beta)}-1} \! \overline{P}^L_\beta (k_\alpha + m \alpha + n_0^L) \right] \\[1.5ex]
        & = 0,
    \end{split}
\end{equation}
implying the absence of exponential divergences, as for the open-string case. Notice that we used the periodicity of the $P^R_\alpha$-function in the right-moving sector and observed the vanishing of the square bracket due to eq.~\eqref{closure} applied to the $P^L_\beta$-function in the left-moving sector.\footnote{Recall that $P_\alpha^R(n)$ has period $\alpha$, while $\overline P_{\beta}^L(n)$ has period $\beta$.} For $\alpha>\beta$, of course, we can simply exchange them above. So, for an exhaustive analysis, we are left with $\alpha=\beta$, in which case one unfortunately cannot generally show a cancellation.  However, if for example only odd $\alpha$s and even $\beta$s appear in the right- and left-moving sectors, respectively, or viceversa, this case is obviously not encountered. This situation is realised for instance in the heterotic $\mathrm{SO}(16) \!\times\! \mathrm{SO}(16)$-theory \cite{Cribiori:2020sct}. For other theories, the cancellations may also happen to take place between different terms in the partition function (i.e. different right-left products cancelling their contributions against each other).  For the time being we defer a completely general analysis.

To summarise, we have proven that misaligned supersymmetry in the state degeneracies leads to a cancellation of the exponential divergence in the one-loop cosmological constant of closed-string theories, if the condition $\alpha \neq \beta$ holds for all right-left products of Kloosterman-like sums. The mathematical structure is exactly the same as for open strings, with minor technical complications only induced by the product of right- and left-moving sectors. The cancellations are also the same as those occurring in the sector-average that is defined in terms of envelope functions, discussed in detail in ref.~\cite{Cribiori:2020sct} using the explicit example of the heterotic $\mathrm{SO}(16) \!\times\! \mathrm{SO}(16)$-theory.

\subsection{Cancellation of polynomial divergences} 
Having cancelled the exponential divergences, the remaining contributions to the one-loop cosmological constant are encoded in the function $g(\tau_2) = \tau_2^{1-D/2} \bigl[ (-1)^{F_{0}} g_{0} + h(\tau_2) \bigr]$, where now the leftover part of the function $h(\tau_2)$ reads
\begin{equation} \label{h final}
    h(\tau_2) = \sum_{m=0}^\infty b_m \tau_2^m,
\end{equation}
with coefficients
\begin{equation}\label{eq:htau2}
    \begin{split}
        b_m & = \sum_{\alpha \in \Gamma_R} \sum_{\beta \in \Gamma_L}  \!\!\sum_{\ell=1}^{\mathrm{lcm}(\alpha,\beta)}\!\! P^R_\alpha (\ell+n_0^R) \overline{P}^L_\beta (\ell+n_0^L) \\[1.0ex]
        & \times \dfrac{4\pi^2 c^R_2(\alpha) c^L_2(\beta)}{\alpha^{-c^R_1} \beta^{-c^L_1}} \sum_{a=0}^\infty \sum_{b=0}^\infty \biggl[ \dfrac{\pi}{12} \dfrac{c^R_3(\alpha)}{\alpha^2} \biggr]^{c^R_1+a+1} \biggl[ \dfrac{\pi}{12} \dfrac{c^L_3(\beta)}{\beta^2} \biggr]^{c^L_1+b+1} \\
        & \times \dfrac{(2 \pi \gamma_{\alpha \beta})^{a+b}}{a! \, b! \, (c^R_1+a+1)! \, (c^L_1+b+1)!} \, B_{m+a+b+1} \biggl( \dfrac{\ell}{\gamma_{\alpha \beta}} \biggr) \dfrac{(-1)^{m+1} (4 \pi \gamma_{\alpha \beta})^{m}}{(m+a+b+1) \, m!}.
    \end{split}
\end{equation}
The term $(-1)^{F_{0}} g_{0}$ can be computed straightforwardly, whilst the coefficients $b_m$ are very difficult to study analytically.

All in all, this is again reminiscent of the open-string result for $g(\tau_2)$ in eq. (\ref{g3}). In that case, we were able to deduce which coefficients $b_l$ were non-vanishing by using properties of the Dedekind $\eta$-quotient, thanks to Zagier. For the closed-string case, $g(\tau_2)$ is not simply a Dedekind $\eta$-quotient, but the integral over d$\tau_1$ of the product of a Dedekind $\eta$-quotient and the complex conjugate of another Dedekind $\eta$-quotient, so the open-string arguments do not follow.  To make progress, we have to be careful about the fact that the partition function may be composed of several terms $Z_\sigma(\tau, \overline{\tau}) = R_\sigma(\tau) \overline{L}_\sigma(\overline{\tau})$, each modular non-invariant but combining to a modular-invariant sum. In this case, the function to be eventually considered is of the form $g(\tau_2) = \tau_2^{1-D/2} \bigl[ (-1)^{F_{0}} g_{0} + \sum_\sigma h^\sigma(\tau_2) \bigr]$, where each function $h^\sigma(\tau_2) = \sum_{m=0}^\infty b_m^\sigma \tau_2^m$ is a power series as outlined in eqs.~ \eqref{h final}, \eqref{eq:htau2}.

Having restored modular invariance, the one-loop cosmological constant is of course finite and the limit of $g(\tau_2)$ is finite too, according to the Kutasov-Seiberg identity in eq. (\ref{KS}).  In fact, in the derivation of the latter, an asymptotic behaviour analogous to the open-string one in eq. (\ref{asymptotic_M}) can be established by considering the Mellin transform $I(s)$ of $g(\tau_2)/\tau_2$. Following refs.~\cite{ZeidlerZagier, Angelantonj:2010ic, Abel:2021tyt}, one can show the relationship
\begin{equation}
    I(s) \equiv \int_{0}^\infty \de \tau_2 \, \tau_2^{s-1} \dfrac{g(\tau_2)}{\tau_2} = \int_{\mathbb{F}} \dfrac{\de^2 \tau}{\tau_2^2} \, E(\tau,\overline{\tau};s) Z(\tau, \overline{\tau}),
\end{equation}
where $E(\tau,\overline{\tau};s)$ is the non-holomorphic Eisenstein series. One can then invert the Mellin transform to write
\begin{equation}
    \dfrac{g(\tau_2)}{\tau_2} \overset{\tau_2 \sim 0^+}{\sim} \sum_{i} r_i \tau_2^{-s_i},
\end{equation}
where $s_i \in \lbrace s_0,s_a \rbrace$ are the poles of the Mellin transform and $r_i$ are the corresponding residues. On the real axis, $s_0=1$ is the only pole. In the rest of the complex plane, these poles can be seen to be related to the non-trivial zeros of the Riemann $\zeta$-function as $s_{a}=\rho_a/2$, where $\rho_a = 1/2 \pm \I \gamma_a$, for $\gamma_a \in \mathbb{R}$, assuming the Riemann hypothesis to be correct. In the function $g(\tau_2)$, the leading term for $\tau_2 \sim 0^+$ is clearly given by the real pole $s_0=1$, implying the finite limit $\smash{\lim_{\tau_2 \to 0^+} g(\tau_2) = r_0}$. The associated residue can be seen to be $r_0=3I/\pi$, where $\smash{I = \int_{\mathbb{F}} \de^2 \tau \, Z(\tau, \overline{\tau}) / \tau_2^2}$. So, in analogy with the open-string result in eq. (\ref{b-coefficient values}), the conclusion is that
\begin{subequations}
\begin{align}
    & (-1)^{F_0} g_0 + \sum_\sigma b^\sigma_{0} = \sum_\sigma b^\sigma_1 = \dots = \sum_\sigma b^\sigma_{D/2-2} = 0, \\
    & \sum_\sigma b^\sigma_{D/2-1} = r_0.
\end{align}
\end{subequations}

It is again useful to consider a concrete example.  For the $\mathrm{SO}(16) \!\times\! \mathrm{SO}(16)$-theory, discussed in more detail in subsection \ref{ssec:hetSO16}, one term in the partition function is of the form
\begin{equation}
    Z(\tau, \overline{\tau}) = \tau_2^{1-D/2} R(\tau) \overline{L}(\overline{\tau}) = \tau_2^{1-D/2} \dfrac{16 \, \eta^{8}(2 \tau)}{\eta^{16}(\tau)} \dfrac{\bar \eta^{8}(\bar\tau)}{\bar\eta^{16}(2\bar \tau)} \,.
\end{equation}
The holomorphic and anti-holomorphic terms can be expanded as Hardy-Ramanujan-Rade-macher sums with odd $\alpha$ and even $\beta$, respectively. This means the cancellation of the exponentially divergent term takes place as discussed in subsection \ref{ssec:closedexpcancel}. Nevertheless, even with this explicit example, performing the sums in eq.~(\ref{eq:htau2}) is hard due to the several terms involved, and characterising the behaviour near $\tau_2 \sim 0^+$ of $g(\tau_2)$ is hard due to the $\tau_1$-integration.  To prove the absence of polynomial divergences, one should consider the full modular-invariant partition function and apply the Kutasov-Seiberg identity.

\section{Open-string supertraces} \label{sec: supertraces}

Besides the one-loop cosmological constant, supertraces also encode interesting information about the spectrum and finiteness of a given string-theory model. In supersymmetric setups, the supertraces $\smash{\mathrm{Str} \, M^{2 \beta} = \sum_{n} (-1)^{F_n} g_{n} \, M_n^{2\beta}}$ vanish due to the perfect matching between bosonic and fermionic degrees of freedom at each mass level. As soon as the matching is perturbed, though, for instance by a (spontaneous) breaking of supersymmetry in the vacuum, the supertraces are no longer zero, but generically they are of the same order of magnitude as the mass splittings.

In string-based models, there is an infinite number of degrees of freedom, and therefore any deviation from supersymmetry implies that standard supertraces are potentially infinite. In order to tackle this issue, ref.~\cite{Dienes:1995pm} proposes the definition of supertraces of the form
\begin{equation} \label{supertraces}
   \mathrm{Str} \, M^{2 \beta} = \lim_{t \to 0} \sum_{n=0}^\infty (-1)^{F_n} g_{n} \, M_n^{2\beta} \, \e^{-2 \pi t M_n^2 / \mu^2}.
\end{equation}
These reduce to the standard supertraces for a finite number of degrees of freedom, but they are also well-defined quantities for theories with an infinite number of fields, since the exponential of the mass operator plays the role of a natural cut-off. In view of eq.~\eqref{supertraces}, using the Kutasov-Seiberg identity (\ref{KS}), one can express the one-loop cosmological constant of a closed-string theory in even $D$ non-compact dimensions as \cite{Dienes:1995pm}
\begin{equation} \label{one-loop cosmological constant / supertraces}
    \tilde{\Lambda}_D = \dfrac{1}{\kappa_D^2 l_s^2} \dfrac{(-4\pi)^{\frac{D}{2}}}{96 \pi (D/2 - 1)!} \, \mathrm{Str} \, \biggl( \dfrac{\alpha' M^2}{4} \biggr)^{\!\frac{D}{2}-1},
\end{equation}
with all the supertraces of smaller powers of $M^2$ being zero, i.e. $\mathrm{Str} \, M^{0} = \mathrm{Str} \, M^{2} = \dots = \mathrm{Str} \, M^{D-4} = 0$, where $\mu=2/\alpha'$. This is interpreted as a generalisation of the QFT-expression for the one-loop cosmological constant, which is a sum of terms depending on the usual supertraces (see e.g. refs.~\cite{Martin:1997ns, Coleman:1973jx}).

Since $\tilde{\Lambda}_D$ is finite in theories exhibiting misaligned supersymmetry, when eq.~\eqref{one-loop cosmological constant / supertraces} holds misaligned supersymmetry is a sufficient condition to guarantee the finiteness of the supertraces. However, for open strings there is no analogue to the relationship (\ref{one-loop cosmological constant / supertraces}), as there is no Kutasov-Seiberg identity that expresses the one-loop cosmological constant in terms of a simple limit.  In this section, we show that an expression like eq.~(\ref{supertraces}) also makes sense for open strings and how to interpret it.

In accordance with the definitions of eqs.~(\ref{openstringLambda}), (\ref{openstringM}) and (\ref{g-definition}), the one-loop cosmological constant for the theory of a D$p$-brane can be written as
\begin{equation}
    \Lambda_{\mathrm{D}p} = - \dfrac{T_{\mathrm{D}p}}{2 \pi} \int_0^\infty \dfrac{\de t}{(2t)^{\frac{p+3}{2}}} \, g (t),
\end{equation}
with
\begin{equation}
    g(t) = \sum_{n=0}^\infty (-1)^{F_n} g_{n} \, \e^{-2 \pi t n}.
\end{equation}
For tachyon-free theories, the integral can diverge at $t=0$, whereas the limit $t \sim \infty$ is finite thanks to the exponential suppression factor $\e^{-2 \pi t n}$, for $n>0$, and the power-law damping $t^{-(p+3)/2}$, for $n=0$. For masses $M_n^2 = n / \alpha'$, setting $\mu^2=1/\alpha'$, the supertraces defined in eq.~(\ref{supertraces}) read
\begin{equation}
    \begin{split}
    \label{StrMopen}
        \mathrm{Str} \, {M}^{2 \beta} & = \lim_{t \to 0} \, \sum_{n=0}^\infty (-1)^{F_n} g_{n} \, \biggl(\dfrac{n}{\alpha'}\biggr)^{\! \beta} \, \e^{-2 \pi t n} = \lim_{t \to 0} \, \biggl[ \biggl( - \dfrac{1}{2 \pi \alpha'} \dfrac{\de}{\de t}\biggr)^{\! \beta} g(t) \biggr].
    \end{split}
\end{equation}
Because we showed that the exponential divergences of the form $\e^{1/t}$ cancel out and the function $g(t)$ is just a series of non-negative powers of $t$ (see eqs.~(\ref{g2}) and (\ref{g3})), the function $g(t)$ can be expanded in a Taylor series around the point $t=0$ as
\begin{equation}
    g(t) = \sum_{\beta=0}^\infty \dfrac{t^\beta}{\beta!} \Bigl[ \Bigl( \dfrac{\de}{\de t} \Bigr)^{\!\beta} g(t) \Bigr]_{t=0} = \sum_{\beta=0}^\infty \dfrac{t^\beta}{\beta!} \, (- 2 \pi \alpha')^\beta \, \mathrm{Str} \, {M}^{2 \beta}.
\end{equation}
Since the integration over $t \in [\epsilon,\infty[$ gives a finite result for an arbitrary $\epsilon \in \mathbb{R}^+$, the potentially divergent term in the cosmological constant corresponds to the part integrated over $t \in [0, \epsilon[$, and it can be written as
\begin{equation}
    \delta \Lambda_{\mathrm{D}p} = - \dfrac{T_{\mathrm{D}p}}{2 \pi} \int_0^\epsilon \dfrac{\de t}{(2t)^{\frac{p+3}{2}}} \, g (t) = - \dfrac{T_{\mathrm{D}p}}{2 \pi} \int_0^\epsilon \dfrac{\de t}{(2t)^{\frac{p+3}{2}}} \, \sum_{\beta=0}^\infty \dfrac{t^\beta}{\beta!} \, (- 2 \pi \alpha')^\beta \, \mathrm{Str} \, {M}^{2 \beta}.
\end{equation}
For any given $\beta$, the integral is convergent if $\beta > (p + 1)/2$, which means that for the cosmological constant to be convergent one needs to have
\begin{equation}
    \mathrm{Str} \, {M}^{2 \beta} = 0, \qquad \text{for} \qquad \beta = 0, 1, \dots, \dfrac{p+1}{2}.
\end{equation}
For an anti-D$p$-brane on top of an O$p$-plane, the cosmological constant can be seen to be finite up to $p=6$, which requires all supertraces to vanish up to $\beta=7/2$. Indeed, the first non-zero supertrace has been argued to be for $\beta=4$, namely $\mathrm{Str} \, {M}^{8}$ \cite{Cribiori:2020sct}. Comparing the Taylor expansion of $g(t)$ with the power series defined in eq.~(\ref{g3}), we see that
\begin{equation}
    b_l + (-1)^{F_0} g_0 \, \delta_{l0} = \dfrac{(-1)^l}{l!} \, (2 \pi \alpha')^l \, \mathrm{Str} \, M^{2l}.
\end{equation}
So, for an anti-D$p$-brane on an O$p$-plane, our proof above that $b_l + (-1)^{F_0} g_0 \, \delta_{l0} = 0$ for $l\neq 4$, implies in fact that the only non-vanishing supertrace is $\mathrm{Str} \, M^{8} = 4! b_4/(2 \pi \alpha')^4$.  This resembles the closed-string result in 10-dimensional flat space, where the first non-zero supertrace is likewise $\mathrm{Str} \, M^{8}$.

One may wonder whether the region near $t \sim \infty$ can provide additional information about the supertraces. This is not the case due to the peculiar properties of string-theory one-loop partition functions under modular transformations. In fact, well-behaved changes under $\mathrm{S}$-transformations typically relate the regions around $t \sim 0$ and $t \sim \infty$. For instance, the function $M=M(\tau)$ describing the excitations of the anti-D$p$-brane/O$p$-plane theory, i.e. $M(\tau) = - 8 \eta^{16}(\tau) \, \eta^{16}(4 \tau)/ \eta^{40}(2\tau)$, transforms as
\begin{equation} \label{anti-Dp/Op S-transformation}
    M(\tau)  \; \overset{\mathrm{S}}{\to} M \Bigl( - \dfrac{1}{\tau} \Bigr) = \dfrac{2^4}{(-\I \tau)^4} M \Bigl( \dfrac{\tau}{4} \Bigr).
\end{equation}
This is manifested in a duality relating anti-D$p$- and anti-D$(6-p)$-branes, for $p<7$, as is apparent from the identity
\begin{equation}
    - \dfrac{T_{\mathrm{D}p}}{2 \pi} \int_0^\infty \dfrac{\de t}{(2t)^{\frac{p+3}{2}}} \, M (\I t) = - \dfrac{T_{\mathrm{D}p}}{2 \pi} \int_0^\infty \dfrac{\de y}{(2y)^{\frac{(6-p)+3}{2}}} \, M (\I y)
\end{equation}
obtained by a simple change of variable $t=1/(4y)$, which means
\begin{equation}
    l_s^p\Lambda_{\overline{\mathrm{D}p}/\mathrm{O}p} = l_s^{6-p} \Lambda_{\overline{\mathrm{D}(6-p)}/\mathrm{O}(6-p)}.
\end{equation}

Such a condition suggests that the information available in the region near $t \sim \infty$ is equivalent to the information available around $t \sim 0$.  This is of course true in closed-string theories, and the same holds for open-string theories if one assumes that the function $M(\tau)$ transforms as $M(\I/\tau_2) = s^{-1/2} \tau_2^{-c_1} M(\I \tau_2/l)$, for some positive constant $l$. Then the leftover integration over $t \in [\epsilon, \infty[$ reads
\begin{equation}
    \delta' \Lambda = - \dfrac{T_{\mathrm{D}p}}{2 \pi} \int_\epsilon^\infty \dfrac{\de t}{(2t)^{\frac{p+3}{2}}} \, g (t) = - \dfrac{T_{\mathrm{D}p}}{2 \pi} \dfrac{s^{-\frac{1}{2}}}{2^{c_1}} \biggl( \dfrac{l}{4} \biggr)^{\! \frac{p+1}{2}-c_1 \!} \int_0^{\frac{1}{l \epsilon}} \dfrac{\de y \, g (y)}{(2y)^{2+c_1-\frac{p+3}{2}}}.
\end{equation}
The potential divergence now comes from $y \sim 0$. Taking again advantage of the expansion of the function $g(t)$, one infers that the integral is finite so long as $\beta > (2 c_1 - p - 1)/2$. So, an IR-UV duality generally remains in the presence of an $\mathrm{S}$-transformation.

\section{Ten-dimensional non-supersymmetric strings \\ and misaligned supersymmetry} \label{sec: allnonSUSYmodels}

In this section, we review the known consistent, tachyon-free 10-dimensional non-supersym-metric models, i.e.~the heterotic $\mathrm{SO}(16) \!\times\! \mathrm{SO}(16)$-theory, the Sugimoto $\mathrm{USp}(32)$-model and the type $0^\prime$B $\mathrm{SU}(32)$-theory, outlined in section \ref{sec: introduction}, and we argue that indeed they all exhibit the defining features of misaligned supersymmetry in part of their spectra.\footnote{We refer the reader to ref.~\cite{Basile:2021vxh} for a recent review on non-supersymmetric strings and to ref.~\cite{Basile:2021mkd} for an analysis of the interactions between branes in them.} This supports the proposal that misaligned supersymmetry is a generic feature of the non-supersymmetric string landscape. 

We will make use of the $\mathrm{so}(8)$-characters $O_8$, $V_8$, $C_8$ and $S_8$, given in terms of Jacobi $\vartheta$-functions as
\begin{align}
    O_8 & = \frac{\vartheta_3^4 + \vartheta_4^4}{2\eta^4},\\
    V_8 & = \frac{\vartheta_3^4 - \vartheta_4^4}{2\eta^4},\\
    S_8 & = \frac{\vartheta_2^4 + \vartheta_1^4}{2\eta^4},\\
    C_8 & = \frac{\vartheta_2^4 - \vartheta_1^4}{2\eta^4}.
\end{align}
We refer the reader to e.g.~appendix A of ref.~\cite{Cribiori:2020sct} for more details on their properties.

\subsection[Heterotic SO(16)xSO(16)-theory]{Heterotic \texorpdfstring{$\boldsymbol{\mathrm{SO}(16) \!\times\! \mathrm{SO}(16)}$}{$\mathrm{SO}(16) \!\times\! \mathrm{SO}(16)$}-theory}\label{ssec:hetSO16}
The heterotic $\mathrm{SO}(16) \!\times\! \mathrm{SO}(16)$-theory has been discussed in detail in ref.~\cite{Cribiori:2020sct} and it has a misaligned spectrum. This is a 10-dimensional model whose misalignment has to be studied in relation to a closed-string partition function involving the product of right- and left-moving sectors. Its partition function can be described in terms of the two functions\footnote{These functions are denoted as $R_1$ and $L_1$, respectively, in ref.~\cite{Cribiori:2020sct}.}
\begin{align}
    \mu_1(\tau) & = \dfrac{2 S_8}{\eta^8}(\tau) = \dfrac{\vartheta_2^4(\tau)}{\eta^{12}(\tau)} = \dfrac{16 \, \eta^{8}(2 \tau)}{\eta^{16}(\tau)}, \label{mu1} \\
    \nu_1(\tau) & = \dfrac{\vartheta_3^8(\tau) \vartheta_4^8(\tau)}{\eta^{24}(\tau)} = \dfrac{\eta^{8}(\tau)}{\eta^{16}(2\tau)}. \label{nu1}
\end{align}
In both cases, one finds $c_1=4$ and $G(\alpha) \geq 0$, therefore the particular Hardy-Ramanujan-Rademacher-expansion discussed by ref.~\cite{sussman2017rademacher} applies and provides complete knowledge over all of the subleading contributions. In particular,  for any number $\omega \in \mathbb{N}_0$ we find for $\mu_1$ that $n_0=0$ and $c_3(2\omega+2)=0$, which means that only odd $\alpha$s appear in the Hardy-Ramanujan-Rademacher-sum in eq.~\eqref{d(n)} and for those we have $c_2(2\omega+1)=1/16$ and $c_3(2\omega+1)=12$, while for $\nu_1$ one finds $n_0=1$ and $c_3(2\omega+1)=0$ so that only even $\alpha$s contribute with $c_2(2\omega+2)=1$  and $c_3(2\omega+2)=24$.

To be precise, not all terms in the partition function are necessarily of the form of eq.~(\ref{mu1}) or eq.~(\ref{nu1}), but, apart from the rescaling $\tau'=2\tau$ that only amounts to index labelling, they differ at most due to 1/2-shifts as
\begin{align}
    \tilde{\mu}_1(\tau) & = R_1(\tau+1/2) = \dfrac{16 \, \eta^{16}(\tau) \eta^{16}(4\tau)}{\eta^{40}(2\tau)}, \label{tildemu} \\
    \tilde{\nu}_1(\tau) & = L_1(\tau+1/2) = - \dfrac{\eta^{8}(2\tau)}{\eta^{8}(\tau)\eta^{8}(4\tau)}. \label{tildenu}
\end{align}
One finds $c_1=4$ and $G(\alpha) \geq 0$ for $\tilde{\nu}_1$, with $n_0=1$ and $c_3(4\omega+2)=0$, and the relevant values in the Hardy-Ramanujan-Rademacher-expansion in eq.~\eqref{d(n)} are $c_2(2\omega+1)=16$, $c_2(4\omega+4)=1$, $c_3(2\omega+1)=6$ and $c_3(4\omega+4)=24$. On the other hand, the function $\tilde{\mu}_1$ is not amenable to the Hardy-Ramanujan-Rademacher-expansion of ref.~\cite{sussman2017rademacher}, but this does not constitute a problem, since for counting the state degeneracies we can just work with $\mu_1$ and keep track of the signs produced by the shift.

\subsection[Sugimoto USp(32)-model]{Sugimoto \texorpdfstring{$\boldsymbol{\mathrm{USp}(32)}$}{$\mathrm{Usp}(32)$}-model}
To introduce the Sugimoto model, it is worthwhile to review briefly its appearance in string theory. Following ref.~\cite{Angelantonj:2002ct}, for a type IIB theory modded out by an orientifold projection and with $n_+$ D9-branes and $n_-$ anti-D9-branes, the Klein-bottle, annulus and Möbius-strip direct-channel amplitudes read\footnote{With respect to the notation in ref.~\cite{Angelantonj:2002ct}, here the hat on the characters with shifted argument is understood for simplicity.}
\begin{subequations}
\begin{align}
    \mathcal{K} & = \dfrac{1}{2} \int_{0}^\infty \dfrac{\de \tau_2}{\tau_2^6} \dfrac{V_8 - S_8}{\eta^8} [2 \I \tau_2], \label{Klein bottle} \\
    \mathcal{A} & = \dfrac{1}{2} \int_{0}^\infty \dfrac{\de \tau_2}{\tau_2^6} \dfrac{(n_+^2 + n_-^2) (V_8 - S_8) + 2 n_+ n_- (O_8 - C_8)}{\eta^8} \biggl[ \dfrac{\I \tau_2}{2} \biggr], \label{annulus brane/antibrane} \\
    \mathcal{M} & = - \dfrac{1}{2} \int_{0}^\infty \dfrac{\de \tau_2}{\tau_2^6} \dfrac{\epsilon_{\NS} (n_+ + n_-) V_8 - \epsilon_{\R} (n_+ - n_-) S_8}{\eta^8} \biggl[ \dfrac{\I \tau_2}{2} + \dfrac{1}{2} \biggr], \label{Möbius strip brane/antibrane}
\end{align}
\end{subequations}
where $\epsilon_{\NS}, \epsilon_{\R} = \pm 1$ are factors depending on the symmetry property of the matrix representing the orientifold action on the Chan-Paton indices,  i.e.~$\gamma_{ij}^T = \epsilon \gamma_{ij}$, which in turn restricts the gauge group for $n$ branes from $\mathrm{U}(n)$ down to $\mathrm{SO}(n)$ and $\mathrm{USp}(n)$ for $\epsilon=1$ and $\epsilon=-1$, respectively. Via the transformation $\ell = 1/(2 \tau_2)$, $\ell = 2 / \tau_2$ and $\ell = 1/(2 \tau_2)$, respectively, the three transverse-channel amplitudes read
\begin{subequations}
\begin{align}
    \tilde{\mathcal{K}} & = \dfrac{1}{2} \, 2^5 \int_{0}^\infty \de \ell \, \dfrac{V_8 - S_8}{\eta^8} [\I \ell], \label{transverse Klein bottle} \\
    \tilde{\mathcal{A}} & = \dfrac{1}{2} \, 2^{-5} \int_{0}^\infty \de \ell \, \dfrac{(n_+ + n_-)^2 V_8 - (n_+ - n_-)^2 S_8}{\eta^8} [\I \ell], \label{transverse annulus brane/antibrane} \\
    \tilde{\mathcal{M}} & = - \dfrac{1}{2} \, 2 \int_{0}^\infty \de \ell \, \dfrac{\epsilon_{\NS} (n_+ + n_-) V_8 - \epsilon_{\R} (n_+ - n_-) S_8}{\eta^8} \biggl[ \I \ell + \dfrac{1}{2} \biggr]. \label{transverse Möbius strip brane/antibrane}
\end{align}
\end{subequations}
The lack of a cancellation among the constant terms proportional to $V_8$ and $S_8$ signals the presence of an NSNS- and an RR-tadpole, respectively. An NSNS-tadpole signals the presence of a dilaton potential in the effective action proportional to $\e^{\phi}$. This, in itself, is believed not to be a fundamental inconsistency of the theory.\footnote{See refs.~\cite{Fischler:1986ci,Fischler:1986tb} for seminal work in this direction.} On the other hand, an RR-tadpole would indicate the violation of an equation of motion for an RR-form field, and therefore is unacceptable. The absence of tadpoles is guaranteed by the conditions
\begin{subequations}\label{eq:tadpoles}
\begin{align}
    & 2^5 - \epsilon_{\NS} (n_+ + n_-) = 0, \label{eq:NSNSttadpole}\\
    & 2^5 - \epsilon_{\R} (n_+ - n_-) = 0.\label{eq:RRtadpole}
\end{align}
\end{subequations}
A simple solution to both constraints is given by $n_-=0$ and $n_+=32$, with $\epsilon_{\NS}=\epsilon_{\R}=1$. This is type I string theory and it contains a stack of D9-branes generating the gauge group $\mathrm{SO}(32)$. Solutions with both $n_+, n_- \neq 0$ suffer tachyonic instabilities, due to the presence of $O_8$ in the direct-channel. A consistent solution with no D9-branes, i.e. $n_+=0$, is represented by the Sugimoto model, which has $\epsilon_{\NS} = \epsilon_{\R} = -1$ and $n_-=32$. This theory contains anti-D9-branes generating the gauge group $\mathrm{USp}(32)$ and it has an NSNS-tadpole, but no RR-tadpole. Effectively, implementing the Jacobi identity $V_8=S_8$, the Sugimoto model is described by the Möbius-strip amplitude in eq.~(\ref{Möbius strip brane/antibrane}), which can be written as
\begin{equation}
   \mathcal{S} = \dfrac{1}{2} \int_{0}^\infty \dfrac{\de t}{t^6} \dfrac{V_8 + S_8}{\eta^8} \biggl[ \I t + \dfrac{1}{2} \biggr],\label{eq:USp(32)}
\end{equation}
after the change of variable $\tau_2 = 2 t$. The integrand can be analysed by considering it as the restriction to imaginary arguments of the function
\begin{equation}
    S(\tau) = \dfrac{1}{2} \dfrac{V_8 + S_8}{\eta^8} \biggl[ \tau + \dfrac{1}{2} \biggr],
\end{equation}
where the power-term has been ignored. In fact, up to a constant factor, this is the function $\tilde{\mu}_1$. Not unexpectedly, this has exactly the same structure as the open-string theory shown to exhibit misaligned supersymmetry in ref.~\cite{Cribiori:2020sct}, i.e.~an anti-D$p$-brane sitting on top of an O$p$-plane. This is precisely the kind of partition functions discussed in section \ref{sec: openstring-misSUSY}, where the partition function is not amenable to the Hardy-Ramanujan-Rademacher-expansion of ref.~\cite{sussman2017rademacher}, but the shifted-argument function is.
This means that the exponential UV-divergences cancel automatically as in eq.~(\ref{g2}). In the classification of ref.~\cite{Cribiori:2020sct}, this is case 1.(a). Note, however, that there is an IR-divergence. Since the function above is $\tilde{\mu}_1$ as defined in eq.~\eqref{tildemu}, its small-$\tau_2$ expansion can be obtained from eqs.~\eqref{eq:Mopen} and \eqref{eq:Mopenexpansion} and starts with $\tau_2^4$. So, the integral in eq.~\eqref{eq:USp(32)} is IR-divergent. This is due to the uncancelled NSNS-tadpole in eq.~\eqref{eq:NSNSttadpole}. For these codimension-zero sources this tadpole leads to a runaway potential for the dilaton and could be cancelled by a non-trivial dilaton profile, see for example refs.~\cite{Dudas:2000ff, Mourad:2017rrl}.

It is interesting to interpret the physical content of the Sugimoto $\mathrm{USp(32)}$-model. The closed-string sector is the same as the one of the type I theory, and it is supersymmetric. The open-string sector presents misaligned supersymmetry, and this is reflected in the fact that the gauge representations of bosons and fermions follow an alternating misaligned pattern: even-mass level bosons are in symmetric representations and even-mass level fermions are in antisymmetric representations of $\mathrm{USp}(32)$, and vice versa at odd mass levels. This can be seen easily by counting the degrees of freedom stemming from the combination of the $V_8$-terms in the annulus and in the Möbius strip to count the bosons, and the $S_8$-terms to count fermions \cite{Mourad:2017rrl}.

\subsection{Type $\mathbf{0^\prime}$B Strings}

Combining the functions $O_8/(\tau_2^4 \eta^8)$, $V_8/(\tau_2^4 \eta^8)$, $S_8/(\tau_2^4 \eta^8)$ and $C_8/(\tau_2^4 \eta^8)$, i.e.~the elements that appear from the Hilbert traces of superstring oscillations, including $(-1)^F$-projectors, it is possible to identify further modular-invariant theories along with the type II ones.

Requiring the theory to have a single graviton and to always have bosons and fermions to contribute with opposite signs, one finds that, along with the type IIA and type IIB theories, two more exist. These are the so-called type 0A and type 0B theories \cite{Dixon:1986iz, Seiberg:1986by} and, following again ref.~\cite{Angelantonj:2002ct}, their partition functions read
\begin{align}
    Z_{0\mathrm{A}}(\tau, \overline{\tau}) = \dfrac{1}{\tau_2^4} \dfrac{O_8 \overline{O}_8 + V_8 \overline{V}_8 + S_8 \overline{C}_8 + C_8 \overline{S}_8}{\eta^{8} \smash{\overline{\eta}}^{8}} [\tau, \overline{\tau}], \label{type 0A partition function} \\
    Z_{0\mathrm{B}}(\tau, \overline{\tau}) = \dfrac{1}{\tau_2^4} \dfrac{O_8 \overline{O}_8 + V_8 \overline{V}_8 + S_8 \overline{S}_8 + C_8 \overline{C}_8}{\eta^{8} \smash{\overline{\eta}}^{8}} [\tau, \overline{\tau}]. \label{type 0B partition function}
\end{align}
These theories do not have any spacetime fermions and therefore they are not supersymmetric. Furthermore, they also both contain a tachyon, as is apparent due to the presence of the term $O_8 \overline{O}_8$.

Unlike the case of type 0A, where chirality cannot be achieved, an orientifold projection of the type 0B theory reveals the existence of a theory with a chiral spectrum hosting both bosons and fermions. Actually, there exist three possible such projections with chiral spectra \cite{Bianchi:1990yu, Sagnotti:1995ga, Sagnotti:1996qj, Blumenhagen:1999ns}, and only one of them, remarkably, removes the tachyon. For this theory, referred to as type $0^\prime$B theory \cite{Sagnotti:1995ga}, the open-descendant direct-channel amplitudes are
\begin{subequations}
\begin{align}
    \mathcal{K} & = - \dfrac{1}{2} \int_{0}^\infty \dfrac{\de \tau_2}{\tau_2^6} \dfrac{O_8 - V_8 - S_8 + C_8}{\eta^8} [2 \I \tau_2], \label{0'B Klein bottle} \\
    \mathcal{A} & = - \dfrac{1}{2} \int_{0}^\infty \dfrac{\de \tau_2}{\tau_2^6} \dfrac{1}{\eta^8} \bigl[ \begin{aligned}[t] & - 2 \, (n_V n_C + n_O n_S) O_8 - 2 (n_V n_S + n_O n_C) V_8 \\
    & + 2 (n_O n_V + n_S n_C) S_8 + (n_O^2 + n_V^2 + n_S^2 + n_C^2) C_8 \bigr] \biggl[ \dfrac{\I \tau_2}{2} \biggr], \end{aligned} \label{0'B annulus} \\
    \mathcal{M} & = \dfrac{1}{2} \int_{0}^\infty \dfrac{\de \tau_2}{\tau_2^6} \dfrac{(n_O - n_V - n_S + n_C) C_8}{\eta^8} \biggl[ \dfrac{\I \tau_2}{2} + \dfrac{1}{2} \biggr], \label{0'B Möbius strip}
\end{align}
\end{subequations}
where $n_O$, $n_V$, $n_S$ and $n_C$ are non-negative integers that are fixed by consistency conditions, which will shortly be discussed. In the transverse channel, these amplitudes read
\begin{subequations}
\begin{align}
    \tilde{\mathcal{K}} & = - \dfrac{1}{2} \, 2^6 \int_{0}^\infty \de \ell \, \dfrac{C_8}{\eta^8} [\I \ell], \label{transverse 0'B Klein bottle} \\
    \tilde{\mathcal{A}} & = \dfrac{1}{2} \, 2^{-6} \int_{0}^\infty \de \ell \, \dfrac{1}{\eta^8} \bigl[ \begin{aligned}[t] & - (n_O + n_V - n_S - n_C)^2 O_8 + (n_O + n_V + n_S + n_C)^2 V_8 \\
    & + (n_O - n_V + n_S - n_C)^2 S_8 - (n_O - n_V - n_S + n_C)^2 C_8 \bigr] [\I \ell], \end{aligned} \label{transverse 0'B annulus} \\
    \tilde{\mathcal{M}} & = \dfrac{1}{2} \, 2 \int_{0}^\infty \de \ell \, \dfrac{(n_O - n_V - n_S + n_C) C_8}{\eta^8} \biggl[ \I \ell + \dfrac{1}{2} \biggr]. \label{transverse 0'B Möbius strip}
\end{align}
\end{subequations}
Focusing on the consistency conditions stemming from eqs.~\eqref{transverse 0'B Klein bottle}, \eqref{transverse 0'B annulus} and \eqref{transverse 0'B Möbius strip}, one should set the coefficients of the $O_8$- and $S_8$-terms to zero, since they describe boson and fermion contributions with the wrong sign: this pair of conditions reduces to $n_O=n_C$ and $n_V=n_S$. Further, tadpole cancellation requires that the ubiquitous $C_8$-contributions vanish, thus fixing $n_O = 32 + n_V$.  Next, in the direct-channel, the closed-string tachyon in the halved torus amplitude stemming from eq.~(\ref{type 0B partition function}) is removed by the Klein-bottle term in eq.~(\ref{0'B Klein bottle}). To additionally remove the open-string tachyon from the annulus term in eq.~(\ref{0'B annulus}), in view of the tadpole constraints, one has to fix $n_V=n_S=0$, which means $n_O=n_C=32$. Note that these conditions still leave a dilaton tadpole \cite{Sagnotti:1995ga} from the $V_8$-term in the transverse-channel annulus.

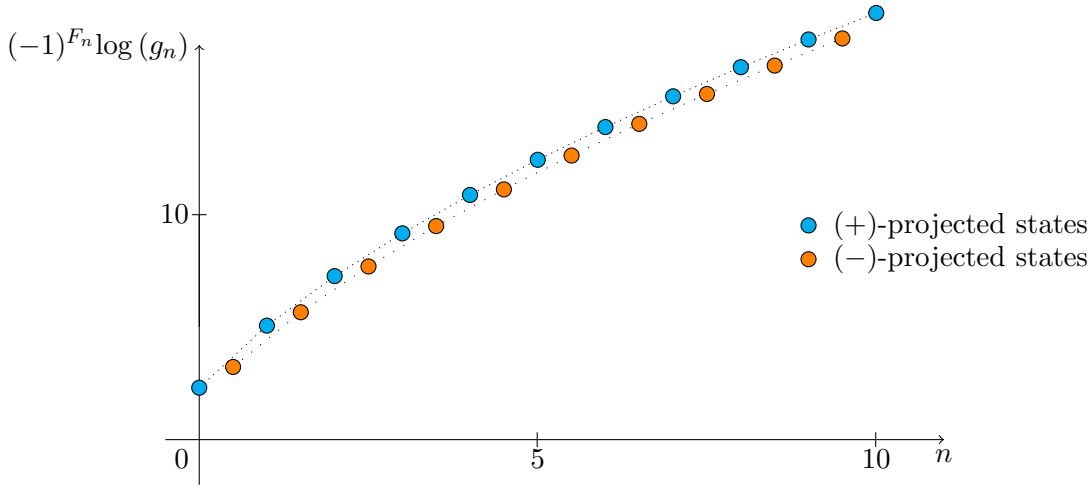
\begin{figure}[ht]
    \centering
    
    \begin{tikzpicture}[xscale=0.90,yscale=0.15,a/.style={draw,circle,minimum size=2mm,inner sep=0pt,outer sep=0pt,black,fill=orange,solid},s/.style={draw,circle,minimum size=2mm,inner sep=0pt,outer sep=0pt,black,fill=cyan,solid}]

    \draw[white] (0,0) -- (17,0);
    
    \draw (-0.5,0) -- (0,0) node[below left]{$0$};
    \draw[-|] (0,0) -- (5,0) node[below]{$5$};
    \draw[-|] (5,0) -- (10,0) node[below]{$10$};
    \draw[->] (10,0) -- (11,0) node[below]{$n$};
    \draw[-|] (0,-4) -- (0,20) node[left]{$10$};
    \draw[->] (0,10) -- (0,35) node[left]{$(-1)^{F_n} \mathrm{log} \, (g_n)$};
    
    \draw[dotted] (0,4.605170186) node[s]{} -- (1,10.11212642) node[s]{} -- (2,14.50426410) node[s]{} -- (3,18.29825816) node[s]{} -- (4,21.70164905) node[s]{} -- (5,24.82208705) node[s]{} -- (6,27.72446980) node[s]{} -- (7,30.45166364) node[s]{} -- (8,33.03382132) node[s]{} -- (9,35.49316036) node[s]{} -- (10,37.84663847) node[s]{};
    
    \draw[loosely dotted] (0.5,6.445719819) node[a]{} -- (1.5,11.29726634) node[a]{} -- (2.5,15.36147946) node[a]{} -- (3.5,18.94167702) node[a]{} -- (4.5,22.19399287) node[a]{} -- (5.5,25.19881869) node[a]{} -- (6.5,28.00925945) node[a]{} -- (7.5,30.66071442) node[a]{} -- (8.5,33.17907428) node[a]{} -- (9.5,35.58361457) node[a]{};
    
    \node[s] at (9,19){};
    \node[right] at (9,19){\, $(+)$-projected states};
    \node[a] at (9,16){};
    \node[right] at (9,16){\, $(-)$-projected states};

    \end{tikzpicture}
    
    \caption{The net number of physical degrees of freedom for the lightest energy levels in the closed-string sector of the type $0^\prime$B theory, defined as $g_n$, at the $n$-th mass level. All states are bosonic, and each point corresponds to states with mass $M_n^2 = 4n / \alpha'$, with $n=0,1/2,1,\dots,10$. There is a clear alternance between states receiving a positive contribution from both the torus and the Klein bottle, i.e. undergoing the `$(+)$-projection', and states receiving a positive contribution from the torus and a negative contribution from the Klein bottle, i.e. undergoing the `$(-)$-projection'.}
    
    \label{type 0'B closed-string plot}
    
\end{figure}

The total one-loop amplitude, proportional to the one-loop cosmological constant, is
\begin{equation} \label{0'B full amplitude}
    \begin{split}
        & \mathcal{T}/2 + \mathcal{K} + \mathcal{A} + \mathcal{M} = \\
        = \, & + \dfrac{1}{2} \int_{\mathbb{F}} \dfrac{\de^2 \tau}{\tau_2^6} \dfrac{\ab O_8 \ab^2 \!+\! \ab V_8 \ab^2 \!+\! \ab S_8 \ab^2 \!+\! \ab C_8 \ab^2}{\ab \eta \ab^{16}}[\tau, \overline{\tau}] - \dfrac{1}{2} \int_{0}^\infty \dfrac{\de \tau_2}{\tau_2^6} \dfrac{O_8 \!-\! V_8 \!-\! S_8 \!+\! C_8}{\eta^8} [2 \I \tau_2] \\
        & - \dfrac{1}{2} \int_{0}^\infty \dfrac{\de \tau_2}{\tau_2^6} \dfrac{\bigl[ - 2 \cdot 32^2 \, V_8 + 2 \cdot 32^2 \, C_8 \bigr]}{\eta^8} \biggl[ \dfrac{\I \tau_2}{2} \biggr] + \dfrac{1}{2} \int_{0}^\infty \dfrac{\de \tau_2}{\tau_2^6} \dfrac{2 \cdot 32 \, C_8}{\eta^8} \biggl[ \dfrac{\I \tau_2}{2} + \dfrac{1}{2} \biggr].
    \end{split}
\end{equation}
One can now discuss the presence of misaligned supersymmetry within this amplitude.  Along with the functions $\mu_1$ and $\tilde{\mu}_1$ of eqs.~ (\ref{mu1}) and (\ref{tildemu}) (recalling that $S_8=C_8$), the functions\footnote{These functions are denoted as $-R_2$ and $R_3$, respectively, in ref.~\cite{Cribiori:2020sct}.}
\begin{align}
    \mu_2(\tau) & = \dfrac{O_8 + V_8}{\eta^8}(\tau) = \dfrac{\vartheta_3^4(\tau)}{\eta^{12}(\tau)} = \dfrac{\eta^{8}(\tau)}{\eta^{8}(\tau/2) \eta^{8}(2\tau)}, \label{mu2} \\
    \mu_3(\tau) & = \dfrac{O_8 - V_8}{\eta^8}(\tau) = \dfrac{\vartheta_4^4(\tau)}{\eta^{12}(\tau)} = \dfrac{\eta^{8}(\tau/2)}{\eta^{16}(\tau)} \label{mu3}
\end{align}
also appear.  As in ref.~\cite{Cribiori:2020sct}, it is convenient to rescale the variable as $\tau'=2\tau$, obtaining
\begin{align}
    \mu'_2(\tau) & = \mu_2(2\tau) = \dfrac{\eta^{8}(2\tau)}{\eta^{8}(\tau) \eta^{8}(4\tau)}, \\
    \mu'_3(\tau) & = \mu_3(2\tau) = \dfrac{\eta^{8}(\tau)}{\eta^{16}(2\tau)}.
\end{align}
Both functions have $c_1=4$ and $G(\alpha) \geq 0$. For $\mu'_2$ one finds $n_0=1$, $c_2(2\omega+1)=16$, $c_2(4\omega+4)=1$,  $c_3(2\omega+1)=6$ and $c_3(4\omega+4)=24$, with $c_3(2 \, \mathrm{mod} \, 4)=0$. For $\mu'_3$, one has $n_0=1$, $c_2(2\omega)=1$ and  $c_3(2\omega)=24$, with $c_3(2\omega+1)=0$. One should notice the identity
\begin{equation}
    \nu_1(\tau) = \mu'_2(\tau+1/2) = \mu'_3(\tau).
\end{equation}
To sum up, one has to study the amplitude of eq.~(\ref{0'B full amplitude}) term by term, but luckily this is a relatively easy task for most contributions. The open-string sector is analogous to the Sugimoto $\mathrm{USp}(32)$-theory one. On the other hand, a plot representing the total number of closed-string states for the type $0^\prime$B theory is in fig.~\ref{type 0'B closed-string plot}.
\begin{itemize}
    \item The open-string sector exhibits misaligned supersymmetry. The annulus amplitude happens to vanish by the Jacobi identity, so it represents a supersymmetric term. On the other hand, the Möbius-strip term is proportional to $\tilde{\mu}_1(\tau)$, and therefore its exponential divergences cancel out in the same way as for anti-D$p$-branes/O$p$-planes and the Sugimoto model. This is a manifestation of misaligned supersymmetry, and it refers to the so-called case 1.(a) in ref.~\cite{Cribiori:2020sct}. In fact, this open-string sector follows exactly the same pattern as the Sugimoto $\mathrm{USp}(32)$-model.
    \item In the closed-string sector, the spectrum is purely bosonic. Yet, we can interpret it using the perspective of a misalignment. To start, one has to observe that the torus amplitude has a tachyonic term which is only cancelled by the combination with the Klein bottle. This eliminates IR-divergences. Then, UV-divergences can be seen to be absent from the spectrum since the Klein bottle, described by the function $\mu'_3$, undergoes the cancellations discussed in section \ref{sec: openstring-misSUSY}. This corresponds to case 1.(b) in ref.~\cite{Cribiori:2020sct}. Although the physical interpretation of this fact cannot be phrased in terms of bosonic and fermionic oscillations, the mathematics is the same and in fact one can observe the cancellation of the divergences of the form $\e^{1/\tau_2}$ coming from $O_8$ and $V_8$. The correct physical interpretation regards the projection undergone by the bosons of the closed-string sector after the interplay of the torus with the Klein bottle. The oscillation given by the function $-\mu'_3(\tau) = - q^{-1} \bigl[ 1 - 8 q + 36 q^2 - 128 q^3 + O(q,0)^4 \bigr]$ implies an alternating pattern in the spectrum when combined with the halved torus \cite{Mourad:2017rrl}, as pictured in fig.~\ref{type 0'B closed-string plot}.
\end{itemize}

The type $0^\prime$B theory illustrates several important points in the closed-string sector. Bearing in mind that the tachyon in the half-torus is cancelled by the tachyon in the Klein bottle, the remaining integration of the torus amplitude is finite since the UV-region is cut off from the domain thanks to modular invariance. This specific result does not require misaligned supersymmetry,\footnote{We thank A. Faraggi and V. Matyas for discussions related to this point.} but also it does not violate the claim that all non-tachyonic modular-invariant theories are either supersymmetric or misalignedly-supersymmetric \cite{Dienes:1994np}, since in fact this specific amplitude technically contains a tachyon. The presence of the tachyon also prevents one from making use of the Kutasov-Seiberg identity. The tachyon is actually removed due to an orientifold projection, which brings in a Klein-bottle amplitude as well as (misalignedly-supersymmetric) open-string sectors. These observations also appear in ref.~\cite{Israel:2007nj}. An interesting analysis of the open strings appearing in the type $0^\prime$B theory is also in ref.~\cite{Dudas:2000sn}.

\section{Conclusions} \label{sec: conclusions}

In this article, we have investigated the mechanisms by which string theory is capable of giving finite results in the absence of spacetime supersymmetry. Working at one-loop level in perturbation theory, we have reviewed how this is possible due to modular invariance, which plays a role even when broken by the worldsheet boundaries. Then, we have interpreted such a finiteness as a consequence of cancellations between bosonic and fermionic terms in the full infinite tower of string states. Since the spectrum is not supersymmetric, such cancellations have been named `misaligned' (or `asymptotic') supersymmetry in the literature, and in fact in this article we have shown that the analogy with standard supersymmetric scenarios is indeed accurate. More precisely, we have shown that, in the class of models we have been considering, the cancellations induced by misaligned supersymmetry in the sector-averaged number of states also occur directly in physical observables, such as the one-loop cosmological constant.

As an aside, we have given an interpretation of supertraces for open strings, relating them to the series coefficients of the function whose integral gives the one-loop cosmological constant. This is reminiscent of the closed-string results of ref.~\cite{Dienes:1995pm}, where the first non-vanishing supertrace is shown to be proportional to the one-loop cosmological constant. It would be interesting to examine the formulation of the light-fermion conjecture proposed in ref. \cite{Gonzalo:2021fma}, which makes use of standard supertraces, in terms of these string-based supertraces for non-supersymmetric models.

Finally, we have discussed the presence of misaligned supersymmetry in all known 10-dimensional non-supersymmetric string constructions. While the heterotic $\mathrm{SO}(16) \!\times\! \mathrm{SO}(16)$-theory and the Sugimoto $\mathrm{USp}(32)$-model, along with the single anti-D$p$-brane/O$p$-plane theory, clearly exhibit misaligned supersymmetry, the type $0^\prime$B theory is more interesting. Its closed spectrum is purely bosonic and thus it cannot realise misaligned supersymmetry, strictly speaking. However, the Klein bottle, needed to remove the closed-string tachyon, does exhibit a misalignment. Likewise the open-string annulus and M\"obius-strip amplitudes do realise it as conjectured in ref.~\cite{Israel:2007nj}. 

We focused on a generic family of partition functions, which could be simply recast in the form of Dedekind $\eta$-quotients. This includes all known 10-dimensional non-supersymmetric closed- and open-string models, but in principle there may exist other models that require  extensions of our analysis. One may need e.g. to relax the assumptions on the form of the partition functions, or on the parity of the values $\alpha$ denoting successively subleading terms in the Hardy-Ramanujan-Rademacher-sums, for both open and closed strings. 
A particularly interesting future direction to pursue is the analysis of more realistic compactified 4-dimensional non-supersymmetric theories, including non-Abelian gauge groups. Such constructions are attracting significant attention of late, as evidence for supersymmetry in nature remains elusive.  For instance, heterotic string models exhibiting misaligned supersymmetry have recently been analysed in refs.~\cite{Abel:2015oxa, Abel:2017rch, Abel:2017vos, Nibbelink:2015vha, Faraggi:2020wej, Faraggi:2020fwg, Faraggi:2020hpy}. Noteworthy constructions involving open strings are for instance refs.~\cite{Dudas:2000sn,Angelantonj:1998gj, Blumenhagen:1999uy, Blumenhagen:1999ad, Blumenhagen:1999ns, Parameswaran:2020ukp, Coudarchet:2021qwc}.

Recently, ref.~\cite{Abel:2021tyt} has discussed the calculation of the one-loop scalar masses in string-theoretic constructions. It would be interesting to analyze the expression of such masses with the tools presented here, in order to see how misaligned supersymmetry acts concretely in observables other than the cosmological constant. Ultimately, one would like to understand to what extent modular invariance, misaligned supersymmetry and the infinite towers of string states can help with the long-standing hierarchy problems in the cosmological constant and Higgs mass, as propounded in ref.~\cite{Dienes:2001se}. A solution to the problem of hierarchies based on the coupling to an infinite tower of states has been proposed in ref.~\cite{Buchmuller:2018eog}:  relating this with the viewpoint of misaligned supersymmetry is an intriguing possibility as well.

\acknowledgments

We would like to thank M. Mertens for his precious help about the asymptotic behaviour of modular forms. We are thankful to I.~Basile, R.~Blumenhagen, K.~Dienes, A.~Faraggi, V.~Matyas, S.~Murthy and B.~Percival for very useful discussions. The work of NC is supported by an FWF grant with the number P 30265 and by the Alexander von Humboldt Foundation. The work of FT is supported by the H. G. Baggs Bequest with the University of Liverpool funding code ULG10047. The work of TW is supported in part by the NSF grant PHY-2013988.

\appendix

\section{Special functions appearing in misaligned supersymmetry} \label{app:specialfunctions}

In this appendix, we collect all the relevant results concerning the special functions that appear in the discussion of misaligned supersymmetry.

\subsection{Asymptotic Expansion for the Dedekind \texorpdfstring{$\boldsymbol{\eta}$}{\eta}-function} \label{subapp:asymptoticeta}
It is instructive to discuss in some detail the derivation of the asymptotic expansion of the Dedekind $\eta$-function written in eq.~(\ref{asymptotic_eta}). There are two ways for doing this: one relies on the modular properties of the function, whilst another is just a result of mathematical analysis.

Based on the definitions in ref.~\cite{ZeidlerZagier}, the notation and the terminology is as follows.
\begin{itemize}
    \item The expression $f(x) = O(g(x); x_0)$ means that there exists a value $M \in \mathbb{R}^+$ such that $\ab f(x) \ab \leq M g(x)$ for any $x$ in a sufficiently small neighbourhood $I_{x_0}$. The expression $f(x) = o(g(x); x_0)$ means that $\lim_{x \to x_0} ( f(x)/g(x) ) = 0$.
    \item The expression $f(x) \overset{x \sim x_0}{\simeq} g(x)$ means that $\lim_{x \to x_0} f(x)/g(x)=1$. The expression $f(x) \overset{x \sim 0}{\sim} \sum_n f_n x^n$ means that $f(x) - \sum_{n = 0}^m f_n x^n = o(x^{m};0)$ for any natural number $m \in \mathbb{N}$.
    \item The function $f:\; \mathbb{R}^+ \to \mathbb{R}$ is a function of rapid decay at the point $x_0$ if $f(x) \overset{x \to x_0}{\to} 0$ faster than any power $(x-x_0)^m$, i.e. $f(x) = o((x-x_0)^m;x_0)$ for any natural number $m \in \mathbb{N}$.
\end{itemize}

The two derivations of the asymptotic behaviour of the function $\eta(\I \tau_2)$ are discussed below. They are reviews of the discussion by Zagier in ref.~\cite{ZeidlerZagier}.

\begin{enumerate}
    \item One can make use of the behaviour of the Dedekind $\eta$-function under the modular group $\mathrm{PSL}_2(\mathbb{Z})$. Under the generating $\mathrm{S}$-transformation $S(\tau) = - 1 / \tau$, the Dedekind $\eta$-function transforms as
    \begin{equation}
        \eta \Bigl( - \dfrac{1}{\tau} \Bigr) = \sqrt{- \I \tau} \, \eta (\tau).
    \end{equation}
    Restricting to the imaginary axis $\tau = \I \tau_2$, one can thus write
    \begin{equation}
    \label{Stransfeta}
        \eta \Bigl( \dfrac{\I}{\tau_2} \Bigr) = \sqrt{\tau_2} \, \eta (\I \tau_2).
    \end{equation}
    The Dedekind $\eta$-function can be written as $\eta(\I \tau_2) = \e^{- \frac{\pi \tau_2}{12}} \prod_{n=1}^\infty (1 - \e^{- 2 \pi n \tau_2} )$, so one finds that
    \begin{equation}
        \mathrm{ln} \, \eta (\I \tau_2) = - \dfrac{\pi \tau_2}{12} + \sum_{n=1}^\infty \mathrm{ln} \, (1 - \e^{- 2 \pi n \tau_2} ) = - \dfrac{\pi \tau_2}{12} + O(e^{-2 \pi \tau_2}; + \infty).
    \end{equation} 
    The magnitude of the subleading terms stems from the Taylor-Maclaurin expansion $\mathrm{ln} \, (1 + x) = O(x;0)$: one finds $\mathrm{ln} \, (1 - \e^{- 2 \pi t} ) = O(e^{-2 \pi t}; + \infty)$. So, combining the $\mathrm{S}$-transformation relation of eq.~\eqref{Stransfeta} and the limit as $1/\tau_2 \sim \infty$, one finds
    \begin{equation}
    \label{logeta}
        \mathrm{ln} \, \eta(\I \tau_2) = - \dfrac{1}{2} \, \mathrm{ln} \, \tau_2 - \dfrac{\pi}{12 \tau_2} + O(e^{-2 \pi/\tau_2}; 0).
    \end{equation}
    This confirms the expansion in eq.~(\ref{asymptotic_eta}) and quantifies the magnitude of the subleading terms. The asymptotic behaviour is a direct consequence of the modular properties of the Dedekind $\eta$-function.
    \item One can make use of tools from mathematical analysis. One can prove the following theorem. \\
    
    \emph{Theorem.} Let the function $g = g(x)$ be defined as
    \begin{equation}
        g(x) = \sum_{m=1}^\infty f(m x),
    \end{equation}
    where $f$ is a smooth function on the positive real line with the following properties:
    \begin{itemize}
        \item at the origin, $f$ has the asymptotic development
        \begin{equation}
            f(x) \overset{x \sim 0}{\sim} b \, \mathrm{ln} \, \dfrac{1}{x} + \sum_{n=0}^\infty f_n x^n;
        \end{equation}
        \item at infinity, $f$ and all of its derivatives are of rapid decay.
    \end{itemize}
    Further, let the definite integral of $f$ be
    \begin{equation}
        I_f = \int_0^\infty \de x \; f(x).
    \end{equation}
    Then, the function $g=g(x)$ at the origin has the asymptotic development
    \begin{equation}
        g(x) \overset{x \sim 0}{\sim} \dfrac{I_f}{x} - \dfrac{b}{2} \, \mathrm{ln} \, \dfrac{2\pi}{x} + \sum_{n=0}^\infty (-1)^n \dfrac{f_n B_{n+1}}{n+1} x^n.
    \end{equation}
    We refer the reader to ref.~\cite{ZeidlerZagier}, section 6.7.4, for details on the proof. This theorem is enough to determine the asymptotic behaviour of the Dedekind $\eta$-function. Let the function $f$ be
    \begin{equation}
        f(x) = \mathrm{ln} \, \bigl( 1 - \e^{-x} \bigr).
    \end{equation}
     This has the asymptotic expansion\footnote{To see this, one can expand the derivative as
    \begin{equation}
        \dfrac{\de f}{\de x} (x) = \dfrac{1}{x} \dfrac{x}{\e^{x} - 1} = \dfrac{1}{x} +  \sum_{n=0}^\infty \dfrac{B_{n+1}}{(n+1)!} x^n
    \end{equation}
    and integrate it to
    \begin{equation}
        f(x;c) = \mathrm{ln} \, x + \sum_{n=0}^\infty \dfrac{B_{n+1}}{(n+1) (n+1)!} x^{n+1} + c.
    \end{equation}
    By requiring that $f(1) = f(1; c)$, for instance, one finds $c = 0$.} and the definite integral
    \begin{align}
        & f(x) \overset{x \sim 0}{\sim} \mathrm{ln} \, x + \sum_{n=1}^\infty \dfrac{B_n}{n \cdot n!} x^n, \\
        & I_f = - \dfrac{\pi^2}{6}.
    \end{align}
    So, one can apply the theorem with $b=-1$, $f_0=0$, $f_{n} = \frac{B_n}{n\cdot n!}$ for $n\geq 1$, and write
    \begin{equation}
       \sum_{m=1}^\infty f(m x) \overset{x \sim 0}{\sim} - \dfrac{\pi^2}{6 x} + \dfrac{1}{2} \, \mathrm{ln} \, \dfrac{2 \pi}{x} + \sum_{n=1}^\infty (-1)^n \dfrac{B_n B_{n+1}}{n \cdot (n+1)!} x^n = - \dfrac{\pi^2}{6 x} - \dfrac{1}{2} \, \mathrm{ln} \, \dfrac{x}{2 \pi} + \dfrac{x}{24}.
    \end{equation}
    In particular, there are no powers beyond $x^1$ since all the even Bernoulli numbers vanish beyond $B_2=1/6$, with moreover $B_0=1$ and $B_1=-1/2$. This can be used to write
    \begin{equation}
        \begin{split}
            \mathrm{ln} \, \eta (\I \tau_2) & = - \dfrac{\pi \tau_2}{12} + \sum_{m=1}^\infty \mathrm{ln} \, (1 - \e^{- 2 \pi m \tau_2} ) \overset{\tau_2 \sim 0}{\sim} - \dfrac{\pi}{12 \tau_2} - \dfrac{1}{2} \, \mathrm{ln} \, \tau_2,
        \end{split}
    \end{equation}
in agreement with eq.~\eqref{logeta}.
\end{enumerate}

\subsection{Bernoulli and Euler Polynomials}
It is useful to collect the relevant expressions used in the main text about Bernoulli and Euler polynomials. The main guidance is ref.~\cite{AS}.

For a given real $x$, Bernoulli and Euler polynomials $B_n=B_n(x)$ and $E_n=E_n(x)$, respectively, are defined as the coefficients appearing in the Taylor expansions
\begin{align}
\label{Bgenfunc}
    \dfrac{t \, \e^{x t}}{\e^t - 1} = \sum_{n=0}^\infty B_n(x) \dfrac{t^n}{n!}, \\
    \dfrac{2 \, \e^{x t}}{\e^t + 1} = \sum_{n=0}^\infty E_n(x) \dfrac{t^n}{n!}.
\end{align}
For the variable $1 - x$, one finds
\begin{align}
\label{Bprop1}
    B_n(1-x) & = (-1)^n B_n(x), \\
    \label{Eprop1}
    E_n(1-x) & = (-1)^n E_n(x).
\end{align}
A simple equation relates them to each other for $n>0$, i.e.
\begin{equation}
\label{EtoB}
    E_{n-1}(x) = \dfrac{2^n}{n} \biggl[ B_{n} \Bigl( \dfrac{x+1}{2} \Bigr) - B_{n} \Bigl( \dfrac{x}{2} \Bigr) \biggr].
\end{equation}
Bernoulli numbers are defined as $B_n = B_n(0)$, whilst Euler numbers are $E_n = 2^n E_n(1/2)$, for all $n \in \mathbb{N}$.

\section{Analytic expression of open-string power-series coefficients} \label{app: bl-derivation}
It is possible to express the power series $\Delta g^\pm (\tau_2)$ appearing in eq.~(\ref{Deltag_alpha^pm}) in an explicit way. This requires knowledge of the coefficients $\smash{f_l(k, m^\pm_\alpha(\beta))}$, which can be gained by going back to their original introduction. From the definition of the Bernoulli polynomials (see eqs.~(23.1.1) and (23.1.8) in ref.~\cite{AS} or equivalently eqs.~\eqref{Bgenfunc} and \eqref{Bprop1} with $t= 2 \pi \gamma_\alpha \tau_2$, $x = m^\pm_\alpha(\beta)/\gamma_\alpha$) we find 
\begin{equation}
    \begin{split}
        \dfrac{\e^{2 \pi (\gamma_\alpha - m^\pm_\alpha(\beta)) \tau_2}}{\e^{2 \pi \gamma_\alpha \tau_2} - 1} & = \dfrac{1}{2 \pi \gamma_\alpha \tau_2} \dfrac{2\pi \gamma_\alpha \tau_2 \, \e^{2 \pi \gamma_\alpha \tau_2 \bigl[ 1 - \frac{m^\pm_\alpha(\beta)}{\gamma_\alpha} \bigr]}}{\e^{2 \pi \gamma_\alpha \tau_2} - 1} \\
        & = \sum_{n=0}^\infty (-1)^n B_n \biggl[ \dfrac{m^\pm_\alpha(\beta)}{\gamma_\alpha} \biggr] \dfrac{(2 \pi \gamma_\alpha \tau_2)^{n-1}}{n!},
    \end{split}
\end{equation}
where $B_n(x)$ are the Bernoulli polynomials. In the expansion of eq.~(\ref{taylor}) one finds that the power-series coefficients read 
\begin{equation} \label{f_l}
    f_l(k,m^\pm_\alpha(\beta)) = \dfrac{(-1)^{l+1}}{l!} \dfrac{(2 \pi \gamma_\alpha)^{l+k}}{k+l+1} B_{k+l+1} \biggl[\dfrac{m^\pm_\alpha(\beta)}{\gamma_\alpha}\biggr].
\end{equation}
Therefore, in the functions $\Delta g^\pm (\tau_2)$ of eq.~(\ref{Deltag^pm}), the power-series coefficients are (insert eq.~\eqref{f_l} in eq.~\eqref{Deltag_alpha^pm} and compare it with eq.~\eqref{Deltag^pm})
\begin{equation} \label{b^pm}
        \begin{split}
            b^\pm_l = \dfrac{(-1)^{l+1}}{l!} & \sum_{\alpha \in \Gamma} 2 \pi c_2(\alpha) \alpha^{c_1} \! \sum_{\beta=1}^{\alpha} P_\alpha(\beta) \\
            \times & \sum_{k=0}^\infty \dfrac{\biggl[\dfrac{\pi}{12} \dfrac{c_3(\alpha)}{\alpha^2} \biggr]^{c_1+k+1\!\!}}{k! \, (c_1+k+1)!} \dfrac{(2 \pi \gamma_\alpha)^{k+l}}{k+l+1} B_{k+l+1} \biggl[\dfrac{m^\pm_\alpha(\beta)}{\gamma_\alpha}\biggr].
        \end{split}
\end{equation}
Now, starting from eq.~(\ref{b^pm}), one can write the total coefficient $b_l=b_l^+-b_l^-$ as
\begin{equation}
    \begin{split}
        b_l = \dfrac{(-1)^{l+1}}{l!} & \sum_{\alpha \in \Gamma} \dfrac{\pi \alpha^{l-1} c_2(\alpha)}{(2\pi)^{c_1-l+1}} \sum_{\beta=1}^{\alpha} \sum_{k=0}^\infty \dfrac{\biggl[\dfrac{\pi^2}{6} \dfrac{c_3(\alpha)}{\alpha} \biggr]^{c_1+k+1\!\!}}{k! \, (c_1+k + 1)!} P_\alpha(\beta) \\
        \times & \dfrac{2^{k+l+1}}{k+l+1} \Biggl[ B_{k+l+1} \biggl[\dfrac{m^+_\alpha(\beta)}{2 \alpha}\biggr] - B_{k+l+1} \biggl[\dfrac{m^-_\alpha(\beta)}{2 \alpha}\biggr] \Biggr],
    \end{split}
\end{equation}
where we used also that $\gamma_\alpha = 2\alpha$ in our case.
By plugging in the definition of $\smash{m}^\pm_\alpha(\beta)$ in eq.~(\ref{m_beta}), one can see that the difference of Bernoulli polynomials can be written as
\begin{equation}
    B_{k+l+1}\biggl[\dfrac{m^+_\alpha(\beta)}{2 \alpha}\biggr] - B_{k+l+1}\biggl[\dfrac{m^-_\alpha(\beta)}{2 \alpha}\biggr] = (-1)^\beta \Biggl[ B_{k+l+1}\biggl(\dfrac{\beta}{2 \alpha} + \dfrac{1}{2}\biggr) - B_{k+l+1}\biggl(\dfrac{\beta}{2 \alpha}\biggr) \Biggr].
\end{equation}
For $n>0$, Bernoulli and Euler polynomials are related by the condition (see eq.~(23.1.27) in ref.~\cite{AS} or eq.~\eqref{EtoB} in appendix  \ref{app:specialfunctions})
\begin{equation}
    E_{n-1}(x) = \dfrac{2^n}{n} \biggl[ B_{n} \Bigl( \dfrac{x+1}{2} \Bigr) - B_{n} \Bigl( \dfrac{x}{2} \Bigr) \biggr],
\end{equation}
so setting $x=\beta/\alpha$ one finds
\begin{equation}
    \begin{split}
        b_l & = \dfrac{(-1)^{l+1}}{l!} \sum_{\alpha \in \Gamma} \dfrac{\pi \alpha^{l-1} c_2(\alpha)}{(2\pi)^{c_1-l+1}} \sum_{\beta=1}^{\alpha} \sum_{k=0}^\infty \dfrac{\biggl[\dfrac{\pi^2}{6} \dfrac{c_3(\alpha)}{\alpha} \biggr]^{c_1+k+1\!\!\!\!}}{k! \, (c_1+k + 1)!} \, (-1)^\beta P_\alpha(\beta) E_{k+l} \Bigl( \dfrac{\beta}{\alpha} \Bigr) \\
        & = \dfrac{(-1)^{l}}{l!} \sum_{\alpha \in \Gamma} \dfrac{\pi \alpha^{l-1} c_2(\alpha)}{(2\pi)^{c_1-l+1}} \sum_{r=0}^{\alpha-1} \sum_{k=0}^\infty \dfrac{\biggl[\dfrac{\pi^2}{6} \dfrac{c_3(\alpha)}{\alpha} \biggr]^{c_1+k+1\!\!\!\!}}{k! \, (c_1+k + 1)!} \, (-1)^r P_\alpha(-r) E_{k+l} \Bigl( 1 -\dfrac{r}{\alpha} \Bigr),
    \end{split}
\end{equation}
where the change of variable $\beta = \alpha - r$ has been employed, knowing that $\alpha$ is odd by assumption, and it has been made use of the periodicity condition $P_\alpha(\alpha-r) = P_\alpha(-r)$. Because the Euler polynomials are such that $E_n(1-x)=(-1)^n E_n(x)$ (see eq.~(23.1.8) in ref.~\cite{AS} or eq.~\eqref{Eprop1} in appendix \ref{app:specialfunctions}), one can conclude that the power-series coefficients read
\begin{equation}
    b_l = \dfrac{\pi}{l!} \sum_{\alpha \in \Gamma} \dfrac{\alpha^{l-1} c_2(\alpha)}{(2\pi)^{c_1-l+1}} \sum_{k=0}^\infty \dfrac{\biggl[\dfrac{\pi^2}{6} \dfrac{c_3(\alpha)}{\alpha} \biggr]^{c_1+k+1\!\!\!\!}}{k! \, (c_1+k + 1)!} \, \sum_{r=0}^{\alpha-1} (-1)^{k+r} P_\alpha(-r) E_{k+l} \Bigl( \dfrac{r}{\alpha} \Bigr).
\end{equation}
This is eq.~(\ref{b}) in the main text.

\vspace{20pt}

\bibliographystyle{JHEP}

\bibliography{refs.bib}

\end{document}